\documentclass[prd,nofootinbib,preprint,superscriptaddress]{revtex4-1}
\usepackage{comment}
\usepackage{amsmath, amssymb, amsthm, graphicx, epsfig, fancyhdr,epsfig}
\usepackage{physics}
\usepackage{braket}
\usepackage{tikz-feynman}
\tikzfeynmanset{compat=1.1.0}
\usepackage{hyperref}
\usepackage{tikzsymbols}
\usepackage{graphicx}
\usepackage{subcaption}
\usepackage{float}
\usepackage{xcolor}
\usepackage{ulem}

\newcommand{\be}{\begin{equation}}
\newcommand{\ee}{\end{equation}}
\newcommand{\bea}{\begin{eqnarray}}
\newcommand{\eea}{\end{eqnarray}}

\newcommand{\as}[1]{\textcolor{red}{[Angus: #1]}}

\definecolor{lime}{HTML}{A6CE39}
\DeclareRobustCommand{\orcidicon}{
	\begin{tikzpicture}
	\draw[lime, fill=lime] (0,0) 
	circle [radius=0.2] 
	node[white] {{\fontfamily{qag}\selectfont \tiny ID}};
	\draw[white, fill=white] (-0.0625,0.095) 
	circle [radius=0.007];
	\end{tikzpicture}
	\hspace{-2mm}
}

\foreach \x in {A, ..., Z}{\expandafter\xdef\csname orcid\x\endcsname{\noexpand\href{https://orcid.org/\csname orcidauthor\x\endcsname}
			{\noexpand\orcidicon}}
}

\begin{document}
\title{Reconstructing Early Primordial Black Hole Domination from Gravitational-Wave Backgrounds}

\author{Daniel del-Corral \orcidA{}}
\email{daniel.corral.martinez@uj.edu.pl}
\affiliation{Faculty of Physics, Astronomy and Applied Computer Science, Jagiellonian University,
30-348 Krakow, Poland}

\author{Angus Spalding \orcidB{}}
\email{angus.spalding1@gmail.com}
\affiliation{School of Physics and Astronomy, University of Southampton,
Southampton SO17 1BJ, United Kingdom}

\begin{abstract}
Primordial Black Holes (PBHs) with masses below $\mathcal{O}(10^9)$g occupy an interesting region of parameter space that is largely inaccessible to conventional observations. Despite evaporating before Big Bang Nucleosynthesis (BBN), these PBHs can naturally generate a period of early matter domination in the early Universe. A primordial gravitational-wave background (GWB) provides a window onto this otherwise inaccessible regime, since the modified expansion history leaves a characteristic spectral feature associated with the onset and end of PBH domination. The locations of these characteristic frequencies can be used to reconstruct the underlying PBH parameters, in particular the PBH mass. For a freely propagating GWB, the location of these frequencies additionally allow the initial PBH abundance to be determined. We derive simple numerical relations that map these characteristic frequencies directly onto the PBH mass and initial abundance. Future GW experiments span a vast frequency range, providing sensitivity to PBH masses from the BBN bound of $\mathcal{O}(10^9)\,\mathrm{g}$ down to the lower bound of $\mathcal{O}(10)\,\mathrm{g}$. We find that the nanohertz signal reported by NANOGrav, if primordial in origin, is already probing PBH masses in the range $4$--$90\,\mathrm{Mg}$. GW observations therefore offer access to a vast region of PBH parameter space that is otherwise beyond the reach of current experiments.
\end{abstract}
\maketitle

\tableofcontents

\section{Introduction}
In 1971, Hawking proposed that black holes could have formed in the early Universe \cite{Hawking:1971ei}, building on the earlier work of Zeldovich and Novikov \cite{Zeldovich:1967lct}. The subsequent development of the formation criteria for primordial black holes (PBHs) \cite{Carr:1974nx,Carr:1975qj}, together with Hawking's discovery that sufficiently light black holes evaporate through quantum emission \cite{Hawking:1974rv}, established PBHs as a rich probe of early-Universe cosmology. Since then, the field has grown enormously, with PBHs studied as dark-matter candidates \cite{Meszaros:1975ef,Chapline:1975ojl,Afshordi:2003zb, Dvali:2025ktz, Dvali:2024hsb}, as seeds for structure and supermassive black-hole formation \cite{Carr:1983,Carr:1984,Freese:1983}, and as probes of physics across a wide range of cosmological epochs. For comprehensive reviews, see Refs.~\cite{Sasaki:2018dmp,Carr:2020,Green:2020jor,Villanueva-Domingo:2021,Carr:2021bzv,Ozsoy:2023ryl,Stamou:2024lqf,Carr:1993aq}.

One particularly interesting possibility is that PBHs were sufficiently abundant to temporarily dominate the energy density of the early Universe \cite{Ghoshal:2023sfa, Bhaumik:2022pil, Wang:2025lti, Mazde:2022sdx, Manno:2026mws}. Since they behave as pressureless matter, their energy density redshifts more slowly than radiation, so an initially subdominant PBH population can come to dominate and generate a primordial early matter-dominated (eMD) era \cite{Papanikolaou:2020qtd,Domenech:2023jve,Domenech:2024wao,Domenech:2020ssp,Papanikolaou:2022chm,Papanikolaou:2024kjb,He:2024luf}. Such a phase can also arise in more complicated post-inflationary histories, for example during preheating \cite{del-Corral:2025fca}.

Such modifications to the expansion history are difficult to probe with conventional cosmological observables. However, they leave a direct imprint on the stochastic GWB, providing a unique observational handle on these otherwise inaccessible epochs \cite{Ghoshal:2026ros, Datta:2025vyu, Roshan:2026yon, Pearce_2024,BARENBOIM2016430,Assadullahi_2009,Alabidi_2013,Kohri_2018,Inomata_2019, Ferreira:2025zeu, Ferreira:2026uzi, Antusch:2024ypp, Spalding:2026pmp, Murayama:2025thw, Mansour:2026sdx, Lozanov:2026dgu, Casey:2026yhc, Barman:2026kab, Allahverdi:2026ske, Barenboim:2026txj, Barenboim:2026zgj, Ghoshal:2023sfa, Cai:2019cdl, Domenech:2020kqm, Ghosh:2025non, Datta:2025yow, Ghoshal:2025ldb, Cheek:2025gvx, Bhaumik:2022zdd}. Stochastic GW backgrounds therefore act as cosmic archivists, preserving information from epochs far beyond the reach of standard probes. We are able to identify two characteristic frequencies in the spectrum of GWs, corresponding to the onset and end of the eMD era~\cite{Ghoshal:2026ros, Spalding:2026pmp}, that can be directly identified with specific parameters of the PBHs.

Many of the most widely studied stochastic GWBs are generated by transient sources and are therefore freely propagating by the time of interest, including primordial tensor modes from inflation \cite{Caprini_2016,Caprini_2018,Caldwell:2018giq,Boyle_2008,Boyle:2005se,Barman:2023ktz}, first-order phase transitions \cite{Caprini_2016,Hindmarsh:2015qta,Crawford:2024nun,Barman:2026kab}, and annihilating domain walls \cite{Vilenkin:1981zs,Vachaspati:1984gt,Blanco_Pillado_2017,Hiramatsu_2010,Barman:2022yos,Barman:2023fad}. This should be contrasted with backgrounds that can remain continuously sourced, such as those from domain walls prior to annihilation \cite{Vilenkin:1981zs,Vilenkin:2000jqa} and cosmic strings \cite{Ghoshal:2025iil,Battye:2026whd,Kibble:1976sj,Vilenkin:1981zs,Vilenkin:2000jqa,Hindmarsh:1994re}. This distinction is important for the reconstruction considered here, since the information that can be extracted from an eMD imprint depends on whether the background is freely propagating or still being sourced. At the same time, stochastic GW searches now span an enormous range of frequencies, with current and next-generation experiments sensitive to conditions in the Universe within its first second \cite{Kite:2020uix,LIGOScientific:2016aoc,LIGOScientific:2016sjg,LIGOScientific:2017bnn,LIGOScientific:2017vox,LIGOScientific:2017ycc,LIGOScientific:2017vwq,Badurina:2021rgt,Graham:2016plp,Graham:2017pmn,Badurina:2019hst,Punturo:2010zz,Hild:2010id,LIGOScientific:2016wof,Reitze:2019iox,Baker:2019nia,AEDGE:2019nxb,Sesana:2019vho,Garcia-Bellido:2021zgu,Carilli:2004nx,Janssen:2014dka,Weltman:2018zrl,EPTA:2015qep,EPTA:2015gke,NANOGRAV:2018hou,Aggarwal:2018mgp,NANOGrav:2020bcs}. The first direct detection of gravitational waves by the LIGO--Virgo collaboration \cite{LIGOScientific:2016aoc,LIGOScientific:2016sjg} established the observational era of GW astronomy, while pulsar timing array collaborations have since reported compelling evidence for a stochastic background at nanohertz frequencies \cite{Carilli:2004nx,Janssen:2014dka,Weltman:2018zrl,EPTA:2015qep,EPTA:2015gke,NANOGrav:2023gor,NANOGrav:2023hvm}.

In this paper, we investigate how an early period of PBH domination can be reconstructed from a primordial GWB. We determine the conditions for PBH domination and numerically track the resulting cosmological evolution, before identifying the characteristic frequencies imprinted on the GW spectrum by the onset and end of the eMD era. We then establish a direct mapping between these observables and the underlying PBH mass and initial abundance, and determine the regions of parameter space accessible to current and future GW experiments.

\textit{This paper is organised as follows:} In section \ref{sec:PBH} we review the formation of PBHs, and in section \ref{sec:conditions-PBH} identify the region of parameter space in which eMD can occur, leading to an observable imprint in any gravitational wave background. In section \ref{sec:GWB} we analyse this imprint and show that it can be used to directly determine the PBH mass and initial abundance in a direct mapping. We also map the parameter space to upcoming gravitational wave experiments, finding the detectable range of PBH masses and initial abundances. Finally, we conclude in section \ref{sec:conclusion}.


\section{Primordial Black Hole Characterization}
\label{sec:PBH}
In what follows, we give a short description of how PBHs are characterized in terms of their mass and abundance. For the latter, we make a distinction depending on whether the collapse occurs during a radiation-dominated (RD) or a matter-dominated (MD) universes. For more exhaustive reviews, see \cite{Carr:2020,Villanueva-Domingo:2021,Carr:2021bzv}.

\subsection{Primordial Black Hole Mass}

Unlike astrophysical black holes formed by the collapse of stars, PBHs are non-baryonic and can span a vast range of masses. This can be understood if one considers that they form with a mass proportional to the horizon mass at the moment of horizon crossing of the fluctuation, 
\begin{equation}
    M_{\text{PBH}}\simeq\gamma M_H\simeq 10^{5}\,{\rm g}\, \left(\frac{\gamma}{0.2}\right)\left(\frac{g_H}{100}\right)^{-1/2}\left(\frac{T}{10^{13}\,{\rm GeV}}\right)^{-2}
\end{equation}
where $\gamma\sim\mathcal{O}(0.1-1)$ is a proportionality factor, and $g_H$ is the effective number of particle degrees of freedom contributing to Hawking radiation. In general, one expects that the fluctuations that produce PBHs span a certain range of wavenumbers $k$, and therefore the corresponding distribution of PBH masses also spans a certain range of masses. However, for a peaked power spectrum of fluctuations, one typically assumes that the majority of the PBHs will form around a single mass, the one corresponding to the fluctuations close to the peak of amplification. We call this mass $M_{\text{PBH},f}$.


\subsection{Mass Fraction}

The mass fraction usually determines the abundance of PBHs. For an extended distribution of PBH masses, it is defined as \cite{Sasaki:2018dmp,Ozsoy:2023ryl}
\begin{equation}
    \beta\equiv\frac{\dd\Omega_{\text{PBH}}}{\dd\ln M_{\text{PBH}}},
\end{equation}
where $\Omega_{\text{PBH}}=\rho_{\text{PBH}}/\rho_\text{tot}$ represents the fractional energy density of the universe in the form of PBHs, and the expression is to be evaluated at the moment of formation of the PBHs. Since PBHs are also thought to explain part of the dark matter abundance (or the totality in certain mass windows), it is common to express their abundance relative to the dark matter abundance as follows
\begin{equation}
    f\equiv\frac{\Omega_{\text{PBH}}}{\Omega_{\text{DM}}},
\end{equation}
where $\Omega_{\text{PBH}}$ and $\Omega_{\text{DM}}$ are the PBH and dark matter fractional energy densities. Using this, one can relate both estimations as follows \cite{Sasaki:2018dmp,Ozsoy:2023ryl}
\begin{equation}\label{eq:beta-DM}
    \beta=\left(\frac{H_{i}}{H_f}\right)^2\left(\frac{a_f}{a_i}\right)^{-3}\Omega_{\text{DM}}f_{\text{PBH}},
\end{equation}
where the subscript $f$ means evaluation at formation time and the subscript $i$ means evaluation at present time. We can use this relation to constrain the abundance of black holes so that it does not exceed the dark matter fraction. Thus, setting $f_{\text{PBH}}=1$ and using $M_{\text{PBH}}\simeq\gamma M_H$ we get \cite{Sasaki:2018dmp,Ozsoy:2023ryl}
\begin{equation}\label{eq:DM-constraint}
    \beta\lesssim1.33\times10^{-9}\left(\frac{\gamma}{0.2}\right)^{-1/2}\left(\frac{g_{*,f}}{106.75}\right)^{-1/12}\left(\frac{M_{\text{PBH}}}{M_{\odot}}\right)^{1/2},
\end{equation}
where $g_{*,f}$ are the relativistic degrees of freedom at formation time. The mass fraction can be computed in terms of the amplitude of the fluctuations entering the horizon. However, at CMB scales, the amplitude of the fluctuations is constrained to be $\mathcal{O}(10^{-5})$, which is too low to produce a significant abundance of PBHs. Thus, one typically needs either a power spectrum with a blue tilt that increases the power towards smaller scales or a peak-type feature in the power spectrum. See \cite{Sasaki:2018dmp,Carr:2020,Green:2020jor,Villanueva-Domingo:2021,Carr:2021bzv,Ozsoy:2023ryl,Stamou:2024lqf,Carr:1993aq} for reviews of these mechanisms. To keep the analysis as general as possible, we ignore the mechanism that originated the fluctuations and focus just on their post-inflationary evolution and domination.

For a RD universe, Carr \cite{Carr:1975qj} systematically analyzed how density perturbations in the early universe could give rise to small black holes, showing that if the overdensity $\delta$ at the moment of horizon crossing in a region exceeded a critical threshold $\delta_c$, the gravity can counteract the pressure gradients and the whole region would collapse into a black hole rather than expanding with the universe. This idea is based on Jeans' length argument, which laid the foundation for the concept of PBHs, and it is still nowadays the standard criterion used in the literature. The threshold $\delta_c$ is an equation-of-state-dependent parameter, as it carries the information about the pressure that the overdensity must overcome. The original estimation by Carr \cite{Carr:1975qj} was $\delta_c\simeq w$, and later, Harada \textit{et al.} \cite{Harada:2013epa} refined it by deriving a new analytical formula based on more solid physical arguments, given by:
\begin{equation}\label{eq:Harada-criterion}
    \delta_c = \frac{3(1+w)}{5+3w} \sin^2{\left(\frac{\pi\sqrt{w}}{1+3w}\right)},
\end{equation}
In good agreement with Carr's criterion. Other refinements to $\delta_c$ can be found in \cite{Niemeyer:1997mt,Niemeyer:1999ak,Musco:2004ak,Musco:2008hv,Polnarev:2006aa,Nakama:2013ica,Musco:2018rwt,Escriva:2019phb}. The standard and most used formalism to relate the mass fraction with the density perturbations during RD is the Press-Schechter (PS) formalism, originally derived in \cite{Press:1973iz}. It is based on the assumption that the smoothed density contrast $\delta_R$ over a scale $R$ follows a Gaussian distribution of variance $\sigma_R$. The expression for $\beta$ is given by \cite{Harada:2013epa,Green:2020jor}
\begin{equation}\label{eq:PS}
    \beta=2\int_{\delta_c}^{\infty}P(\delta_R)\dd\delta_R\simeq\text{erfc}\left(\frac{\delta_c}{\sqrt{2}\sigma_R}\right)\simeq\text{erfc}\left(\frac{\delta_c}{\sqrt{2\mathcal{P}_{\mathcal{\delta}}(k)}}\right),
\end{equation}
where in the last step we have approximated $\sigma_R^2\simeq\mathcal{P}_{\delta}(k)$, which is valid for a peaked power spectrum of density perturbations, see \cite{Ozsoy:2023ryl} for further details. \footnote{The factor of 2 in eqn.~\eqref{eq:PS} comes from the original Press-Schechter estimation, although some recent research \cite{Kushwaha:2025zpz} found that it is not required when strictly applied to PBH formation.} We can define this spectrum in the usual way as
\begin{equation}
    \mathcal{P}_{\delta}(k)=\frac{k^3}{2\pi^2}|\mathcal{\delta}_{k}|^2,
\end{equation}
where the wavenumber $k$ indicates that we are working in the Fourier space. In general, the quantity that we measure is the spectrum of curvature perturbations, $\mathcal{P}_{\mathcal{R}}(k)$, which can be related to $\mathcal{P}_{\delta}(k)$ \cite{del-Corral:2023apl,Martin:2019nuw,Mukhanov:1990me,Mukhanov:2005sc,Young:2014ana}. Several other formalisms can be used, such as the compaction function approach \cite{Shibata:1999zs}, where full general-relativistic numerical codes are employed, or the peaks theory \cite{Green:2004wb}, where PBHs form from local maxima of the perturbation field. 

As it became evident from Carr's criterion, $\delta_c\simeq w=0$, during a MD phase, the threshold vanishes, indicating the breakdown of this concept for MD scenarios. Since the absence of pressure during this phase allows any fluctuation to collapse into a PBH, one needs to consider other effects that regulate the collapse. Khlopov and Polnarev (KP) were the first ones to study these scenarios \cite{Khlopov:1980mg,Khlopov:1982ef,Polnarev:1985btg,Khlopov:2008qy,Polnarev:1981}, and nowadays it remains the only formalism specifically designed PBH formation during MD phases. Recently, Harada \textit{et al} refined the KP formalism for the anisotropy \cite{Harada:2016mhb}, spin \cite{Harada:2017}, inhomogeneity \cite{Kokubu:2018fxy}, and velocity dispersion \cite{Harada:2022xjp} effects, which are related to the statistical properties of the fluctuations. For instance, considering just the anisotropy effect, the mass fraction can be estimated as
\begin{equation}
    \beta\simeq0.0556\,\sigma^5_R\simeq0.0556\,\mathcal{P}_{\delta}^{5/2}(k),
\end{equation}
for $\sigma\ll0.01$. Although some effort has been put into this direction \cite{Jedamzik:2010dq,Martin:2019nuw,Martin:2020fgl,del-Corral:2023apl,del-Corral:2025lcp}, this scenario remains largely unexplored and poorly understood.

\section{Conditions for Early PBH domination}\label{sec:conditions-PBH}

In this section, we establish the conditions under which PBHs come to dominate the energy density of the Universe, giving rise to an eMD phase. We first provide a numerical treatment of the background evolution, tracking in particular the effective equation of state throughout the PBH-dominated era and its transition back to RD. This evolution is especially relevant for the GWB, whose propagation after horizon crossing depends sensitively on the background equation of state. The resulting dynamics will therefore provide the basis for the GW analysis presented in Sec.~\ref{sec:GWB}. We then compile the latest observational constraints on the PBH parameter space and compare them with the region accessible to primordial GW observations.

\subsection{Numerics}
\begin{figure}[h!]
\centering
\includegraphics[width=1\linewidth]{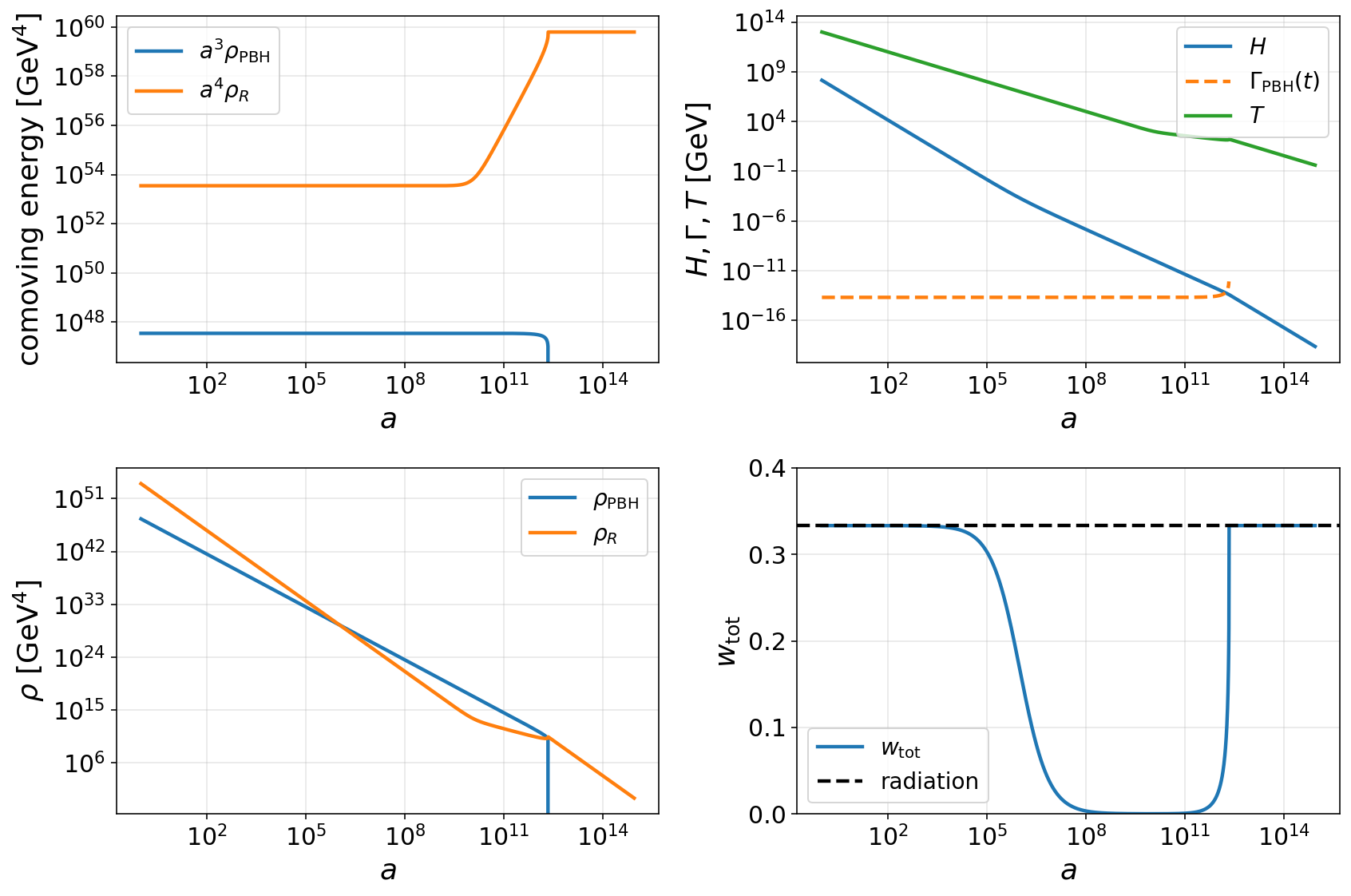} 
\caption{\it Cosmological evolution for the benchmark point $\beta_i=10^{-6}$ and
$T_i=10^{13}\,\mathrm{GeV}$, with $g_*=g_H=106.75$ and $\gamma=0.2$.
The upper-left panel shows the comoving PBH and radiation energy densities,
$a^3\rho_{\rm PBH}$ and $a^4\rho_R$, respectively. The upper-right panel
shows the Hubble rate $H$, the PBH evaporation rate $\Gamma_{\rm PBH}$,
and the radiation temperature $T$. The lower-left panel displays the
physical PBH and radiation energy densities, while the lower-right panel
shows the total equation-of-state parameter $w_{\rm tot}$. The Universe
initially undergoes RD, followed by an intermediate
PBH-dominated era in which $w_{\rm tot}\simeq0$. Once evaporation becomes
efficient, the PBH energy density is transferred to radiation and the
Universe returns to RD, $w_{\rm tot}\simeq1/3$. Further discussion is given in the text.}
    \label{fig:EMD bench}
\end{figure}

\begin{figure}[h!]
\centering
\includegraphics[width=1\linewidth]{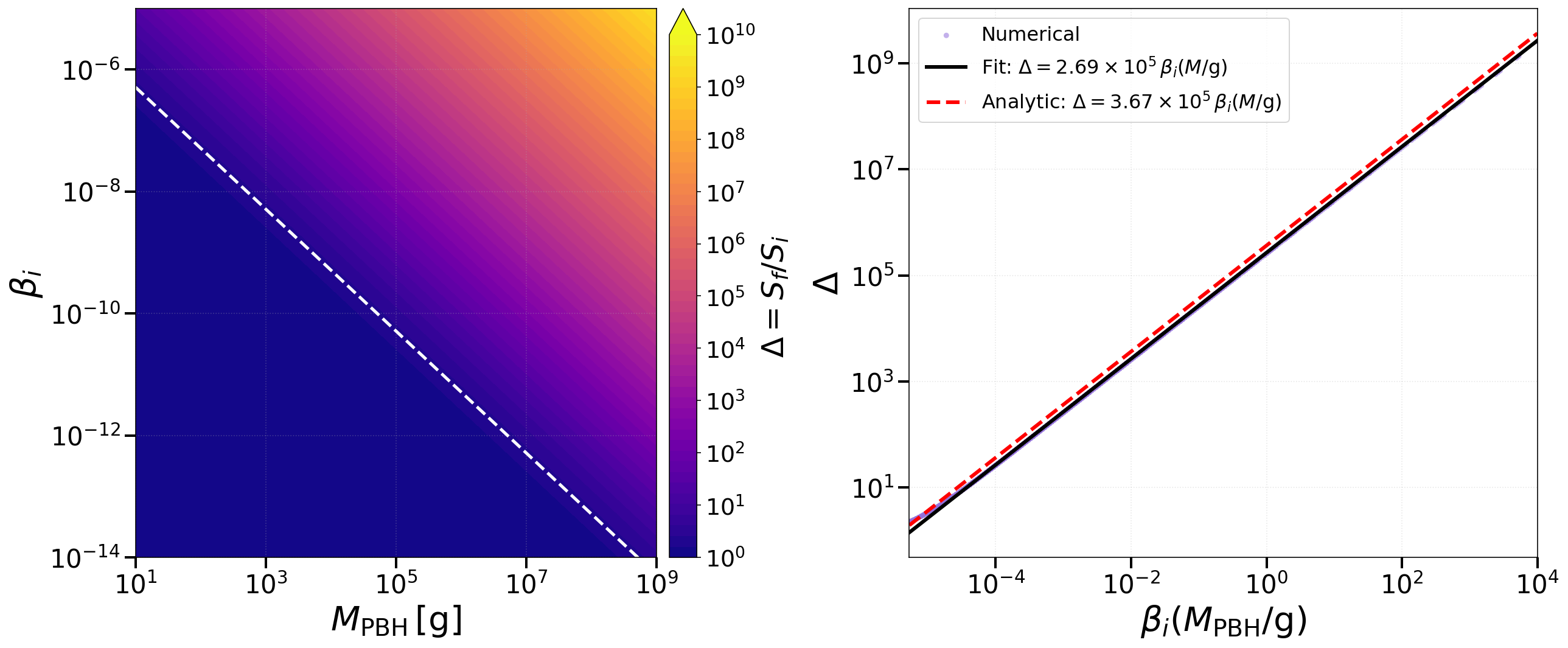} 
\caption{\it Entropy injection from PBH evaporation. \textbf{Left panel:} shows the comoving entropy injection factor, $\Delta=S_f/S_i$, across the $(M_{\rm PBH},\beta_i)$ parameter space obtained from the full numerical evolution. The white dashed line denotes the boundary above which the PBHs undergo a period of eMD. Below this line the dilution factor is unity by definition. \textbf{Right panel:} shows the numerical results within the PBH-dominated region as a function of the combination $\beta_i(M_{\rm PBH}/{\rm g})$ for points satisfying $\beta_i>\beta_{\rm crit}$. The black solid line denotes the one-parameter numerical fit, while the red dashed line shows the analytic estimate obtained in the sudden-evaporation approximation.}
    \label{fig:entropy}
\end{figure}

The cosmological evolution of a population of evaporating PBHs can be described by the standard set of coupled continuity equations for the PBH and radiation energy densities. Treating the PBHs as a pressureless matter component and Hawking evaporation as an effective transfer of energy into relativistic radiation, the background evolution is governed by
\begin{align}
    \dot{\rho}_{\rm PBH}
    + 3H\rho_{\rm PBH}
    &=
    -\Gamma_{\rm PBH}\rho_{\rm PBH},
    \label{eq:pbh_boltzmann}
    \\
    \dot{\rho}_{R}
    + 4H\rho_{R}
    &=
    \Gamma_{\rm PBH}\rho_{\rm PBH},
    \label{eq:rad_boltzmann}
\end{align}
where $H$ is the Hubble parameter, determined by the Friedmann equation,
\begin{equation}
    H^2
    =
    \frac{8\pi}{3M_{\rm pl}^2}
    \left(
        \rho_{\rm PBH}+\rho_R
    \right),
\end{equation}
where $M_{\rm pl}$ denotes the unreduced Planck mass. We model Hawking evaporation as an effective decay of the PBH energy density into radiation, assuming that the emitted particles are relativistic and rapidly thermalize. The corresponding evaporation rate is determined by the Hawking mass-loss rate,
\begin{equation}
    \Gamma_{\rm PBH}
    \equiv
    -\frac{\dot M}{M}
    =
    \frac{\alpha}{M^3},
    \qquad
    \frac{dM}{dt}
    =
    -\frac{\alpha}{M^2},
    \qquad
    \alpha
    =
    \frac{g_H}{30720\pi}M_{\rm pl}^4\ .
\end{equation}
For constant $g_H$, the PBH mass evolves as
\begin{equation}
    M(t)
    =
    M_i
    \left(1-\frac{t}{\tau}\right)^{1/3},
    \qquad
    \tau =
    \frac{M_i^3}{3\alpha}.
\end{equation}
As we showed above, the initial PBH mass is set by the Hubble scale at the time of horizon re-entry. The cosmological evolution is therefore fully specified by two independent parameters: the PBH mass and their initial abundance. We illustrate this evolution for the benchmark choice
$\beta_i=10^{-6}$ and $T_i=10^{13}\,\mathrm{GeV}$, corresponding to
$M_{\rm PBH}\simeq1.9\times10^{5}\,\mathrm{g}$  in Fig.~\ref{fig:EMD bench}.
\\
The benchmark clearly exhibits the characteristic sequence of RD, PBH domination, and the subsequent restoration of RD following PBH evaporation. During the PBH-dominated epoch,
the total equation of state approaches $w_{\rm tot}\simeq0$, while the
comoving radiation energy density increases as Hawking evaporation injects
energy into the thermal bath. At the same time, the rapid growth of
$\Gamma_{\rm PBH}$ as the PBH mass decreases causes evaporation to become
increasingly efficient, culminating in the disappearance of the PBH
component. The associated energy injection temporarily modifies the
temperature evolution relative to the standard adiabatic scaling
$T\propto a^{-1}$ before the usual RD evolution is
recovered. \\
Having illustrated the cosmological evolution for a representative benchmark point, we now scan over the PBH parameter space to determine the regions in which PBHs significantly modify the expansion history and thermal evolution of the early Universe.
\begin{figure}[h!]
\centering
\includegraphics[width=0.7\linewidth]{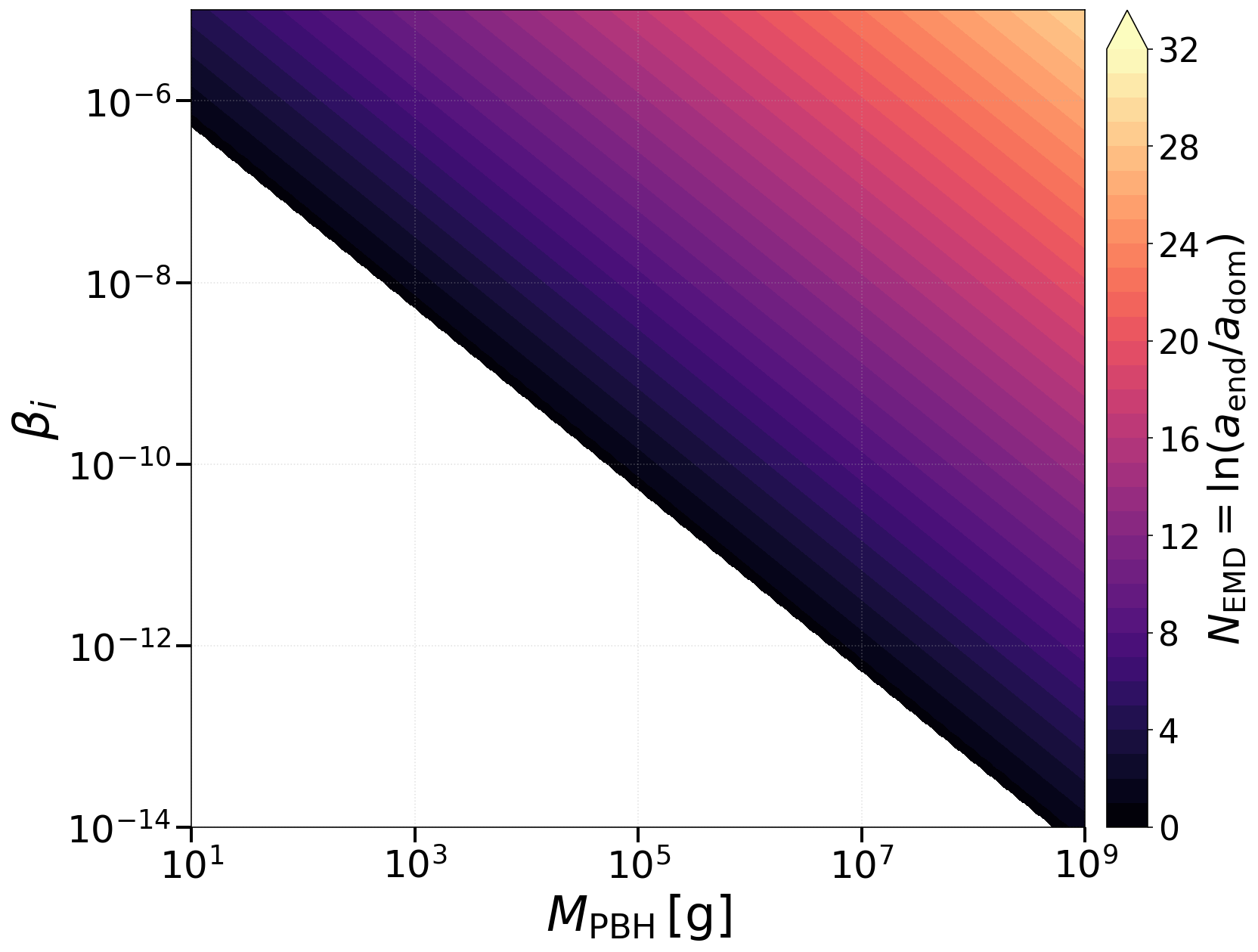} 
\caption{\it Duration of the eMD epoch induced by primordial black
holes in the $(M_{\rm PBH},\beta_i)$ plane. The colour scale shows the
number of $e$-folds of PBH domination,
$N_{\rm EMD}=\ln(a_{\rm end}/a_{\rm dom})$, where $a_{\rm dom}$ and $a_{\rm end}$ are defined by the first and second crossings of $\rho_{\rm PBH}=\rho_R$, respectively. Larger initial PBH abundances and larger PBH masses lead to a longer period of PBH domination, since the former causes PBHs to dominate earlier while the latter increases their lifetime. The white region corresponds to parameter values for which
$\rho_{\rm PBH}$ never exceeds $\rho_R$ before evaporation, and hence no PBH-dominated epoch occurs.}
    \label{fig:EMD scan}
\end{figure}
We scan over the PBH mass and initial abundance and determine the resulting duration of the eMD phase. The number of \(e\)-folds of PBH domination across this parameter space is shown in Fig.~\ref{fig:EMD scan}.\\
From the full numerical evolution, we find that the condition for an early PBH-dominated era is
\begin{equation}
    \beta_{\rm crit}^{\rm num}
    \simeq
    5.11\times10^{-6}
    \left(\frac{M_{\rm PBH}}{\rm g}\right)^{-1}\ .
    \label{eq:PBH_dom_condition}
\end{equation}
such that PBH domination occurs for
$\beta_i>\beta_{\rm crit}^{\rm num}(M_{\rm PBH})$. We derive the corresponding analytic approximation in App.~\ref{appendix:PBHdom}, which reproduces the scaling $\beta_{\rm crit}\propto M_{\rm PBH}^{-1}$ and yields a slightly smaller proportionality constant than the full numerical result. This difference in normalisation arises from the approximations entering the analytic treatment, in particular the neglect of continuous radiation injection during evaporation: some evaporation occurs before $\Gamma_{\rm PBH}=H$, leading to the slightly larger numerical bound.\\
The evaporation of the PBHs also injects entropy into the radiation bath. A key quantity controlling the modification of a GWB by an eMD era is the change in comoving entropy associated with the decay of the dominating component. In our case, this entropy injection is sourced by PBH evaporation. Defining the comoving entropy as $S\equiv sa^3$ with $s$ being the entropy density, and assuming $g_*=g_{*s}$ remains constant throughout the evolution, we define the entropy injection factor as
\begin{equation}
\Delta\equiv\frac{S_f}{S_i}=\frac{s_fa_f^3}{s_ia_i^3}
=\left(\frac{\rho_{r,f}a_f^4}
{\rho_{r,i}a_i^4}\right)^{3/4}.
\end{equation}
An analytic approximation, derived in App.~\ref{appendix:entropy} under the assumption of instantaneous PBH evaporation, predicts the scaling $\Delta\propto\beta_i M_{\rm PBH}$. We test this relation using the full numerical evolution and present the resulting parameter scan, together with a one-parameter best fit, in Fig.~\ref{fig:entropy}.

\noindent The numerical results confirm the expected scaling with the single combination $\beta_i M_{\rm PBH}$. Restricting to the region in which the PBHs dominate the energy density, we obtain the one-parameter fit
\begin{equation}
\Delta=2.69\times10^{5}\beta_i \left(\frac{M_{\rm PBH}}{\mathrm{g}}\right), \quad \beta_i>\beta_{\rm crit}
\end{equation}
in excellent agreement with the linear scaling predicted analytically. The numerical normalisation is slightly smaller than the sudden-evaporation estimate in Eq.~\eqref{eq:Delta_analytic_numeric}. As with the domination criteria result this difference arises primarily because Hawking evaporation is continuous rather than instantaneous. In the sudden-evaporation approximation, the entire PBH energy density is converted into radiation at the end of the PBH lifetime, whereas in the full evolution a fraction of the PBH mass is radiated at earlier times. This radiation subsequently redshifts before evaporation is complete, reducing the final comoving entropy relative to the instantaneous-decay approximation.\\
Having determined the evolution of the equation-of-state parameter and the associated entropy injection, we are now equipped to assess the resulting impact of the PBH-dominated era on the GWB.
\subsection{Constraints on the abundance of PBHs}\label{sec:constraints}

\begin{figure}[t]
        \centering
        \includegraphics[width=0.75\linewidth]{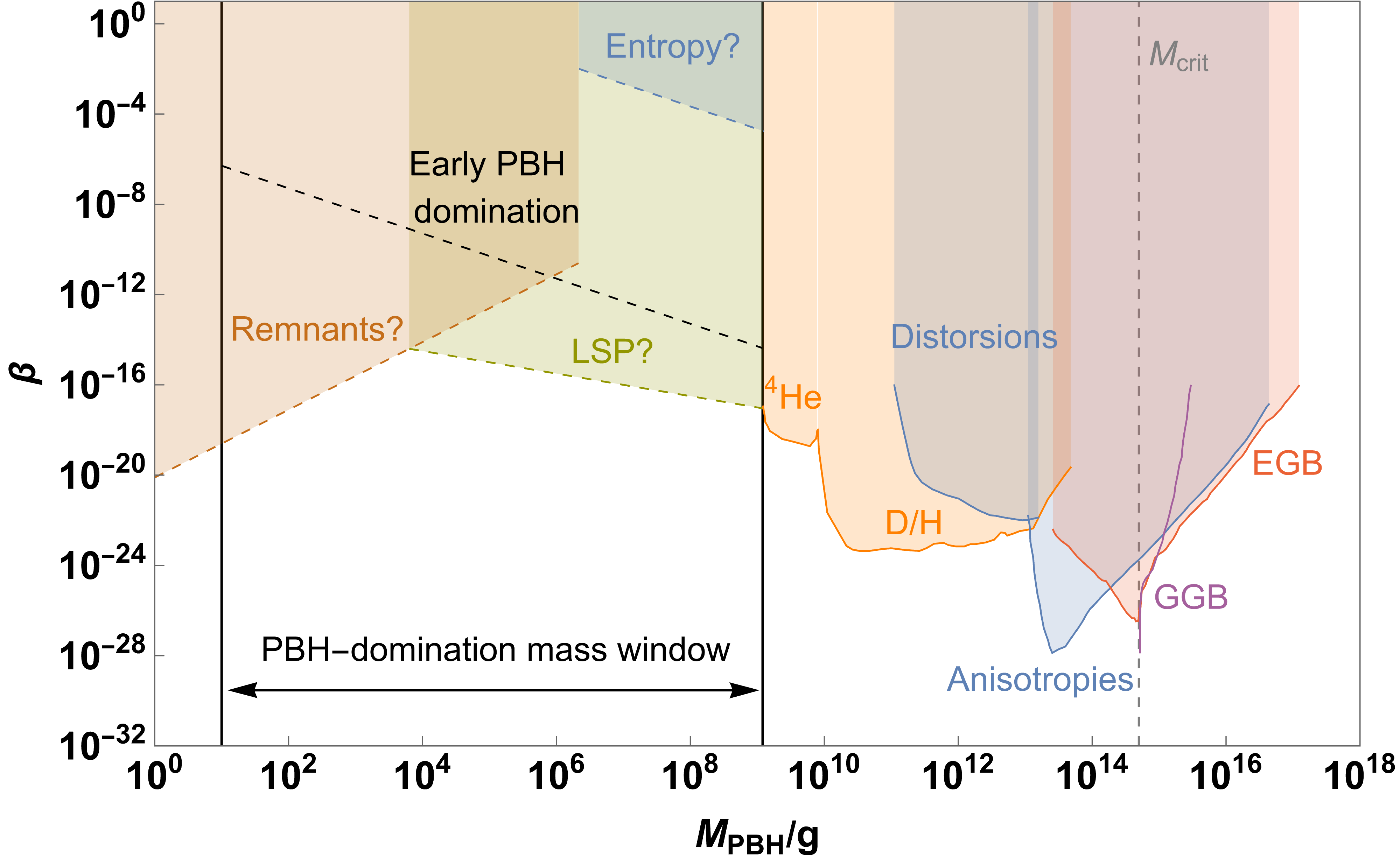}
        \caption{Constraints on evaporating PBHs from Dark matter abundance (DM) \cite{Sasaki:2018dmp,Ozsoy:2023ryl} and Hawking radiation effects. The latter accounts for Planck remnants, LSP emission, CMB effects (entropy, distortions, and anisotropies), BBN effects on the abundance of light elements ($^4$He and D/H), and extragalactic and galactic $\gamma$-ray emission (EGB and GGB, respectively). See \cite{Carr:2020} and references therein for further details. The vertical gray-dashed line corresponds to the critical mass $M_{\text{crit}}$, below which PBHs have already evaporated by the present time. All the data used to construct this figure have been obtained from \cite{Sasaki:2018dmp,Ozsoy:2023ryl,Carr:2020}. The PBH-domination mass window corresponds to the bounds derived in the introduction by requiring a successful reheating temperature. The black dashed line labeled as Early PBH-domination corresponds to the numerical estimate from \eqref{eq:PBH_dom_condition}.}
    \label{fig:constraints}
\end{figure}

As derived in the introduction, when PBHs dominate in the early universe, the reheating is achieved via Hawking radiation. For a successful BBN, this implies that the PBH masses must lie within the range $10g\lesssim M_{\text{PBH}}\lesssim 10^9g$ \cite{Papanikolaou:2020qtd,del-Corral:2025fca,Carr:2020,Carr:2009jm, Boccia:2024nly}. This range is below the current critical mass $M_{\text{crit}}\simeq5\times10^{14}g$, below which PBHs have already evaporated at present. Therefore, the constraints that apply to these PBHs are mainly based on the effects that the emitted particles through Hawking evaporation may have on the surrounding medium, and on the products of the evaporation not overcoming the present abundance of dark matter. For the mass window we are interested in, the main constraints come from:

\begin{itemize}

    \item \textit{Stable Planck-mass remnants.} Hawking evaporation may end on a stable remnant of mass $M_{\text{rem}}=\kappa M_{\text{pl}}$, which behave as non-relativistic matter. Requiring their present density not to exceed the critical density gives \cite{Carr:1994ar}
    \begin{equation}
        \beta(M)<8\times10^{-21}\,\kappa^{-1}\left(\frac{M}{1\,\text{g}}\right)^{3/2}.
    \end{equation}
    Which applies in the following range of masses
    \begin{equation}
        \left(\frac{T_{\rm pl}}{T_R}\right)^2M_{\text{pl}}<M<\kappa^{2/5}\,2.2\times10^{6}\text{g}.
    \end{equation}
    The lower limit excludes PBHs formed before reheating, whose abundance would be exponentially diluted by inflation. The upper limit corresponds to the onset of PBH domination before evaporation. Above this limit, the standard RD relic constraint is no longer applicable: the entropy and baryon-to-photon ratio after evaporation depend on the particles and asymmetries generated by the Hawking emission, so the resulting constraint becomes model-dependent. This bound is conditional on the existence and stability of the remnants, so it should not be taken as a formal bound.

    \item \textit{Stable lightest supersymmetric particles (LSPs).} If the LSP is stable and emitted during the evaporation, its accumulated density must not exceed the observed cold-dark-matter density. The resulting approximate constraint is
    \cite{Lemoine:2000sq}
    \begin{equation}
        \beta(M)\lesssim10^{-18}\left(\frac{M}{10^{11}\,\text{g}}\right)^{-1/2}\left(\frac{m_{\text{LSP}}}{100\,\text{GeV}}\right)^{-1},
    \end{equation}
    which is valid for
    \begin{equation}
        M<10^{11}
        \left(\frac{m_{\text{LSP}}}{100\,\text{GeV}}\right)^{-1}
        \text{g}
    \end{equation}
    so that the PBH becomes sufficiently hot to emit
    particles of mass $m_{\text{LSP}}$. This constraint is also strongly model-dependent because it depends on the existence, mass, stability, and emission rate of the LSP.

    \item \textit{Entropy production.} Photons emitted by sufficiently light PBHs are released early enough to be efficiently thermalized. Their evaporation therefore increases the photon entropy and can dilute a pre-existing baryon asymmetry. Requiring that this dilution does not reduce the baryon-to-photon ratio below its observed value gives the bound
    \begin{equation}
        \beta(M)<10^{-5}\left(\frac{M}{10^{9}\,\mathrm{g}}\right)^{-1},
    \end{equation}
    for $10^{4}\,\mathrm{g}<M<10^{9}\,\mathrm{g}$ \cite{Zeldovich:1977}. This constraint assumes that the baryon asymmetry is generated before PBH evaporation and that its initial value cannot exceed the observed asymmetry by an arbitrarily large factor. If baryogenesis occurs after PBH evaporation, the PBH-generated entropy does not dilute the subsequently produced baryon asymmetry, and this particular constraint is avoided. If baryogenesis occurs during PBH evaporation, the baryon production and entropy injection must instead be treated simultaneously.

\end{itemize}
In Fig.~\ref{fig:constraints}, we show the constraints associated with Hawking radiation effects for the evaporating PBHs ($M_{\text{PBH}}<M_{\text{crit}}$). The vertical black lines determine the PBH-domination mass window, and the vertical dashed line the critical evaporation mass. The constraints affecting PBHs in the range of interest for this article turn out to be particularly model-dependent and grounded on several assumptions, such as the stability or even existence of Planck remnants and LSP particles, and the precise moment of baryogenesis. For this reason, they are shown as dashed lines and with question marks. The rest of the constraints shown in Fig.~\ref{fig:constraints} rely on less speculative effects, see \cite{Carr:2020} and references therein for details on these other constraints. 

\section{Gravitational Wave Background Signatures}
\label{sec:GWB}
Having established the conditions for PBH domination and determined the corresponding cosmological evolution, we now turn to its imprint on the GWB. We first identify the characteristic frequencies associated with the onset and end of the eMD era and determine how these features are encoded in the GW spectrum. We then numerically fit these frequencies in terms of the PBH mass and initial abundance, establishing a one-to-one mapping between observable spectral features and the underlying PBH parameters. Finally, we distinguish between continuously sourced GWBs, for which both transitions may be imprinted in the spectrum, and finite-duration sources, for which the accessible information depends on when the source ceases relative to the PBH-dominated epoch.

\subsection{Gravitational Wave Signatures of Early Matter Domination}
\begin{figure}[h!]
\centering
\includegraphics[width=1\linewidth]{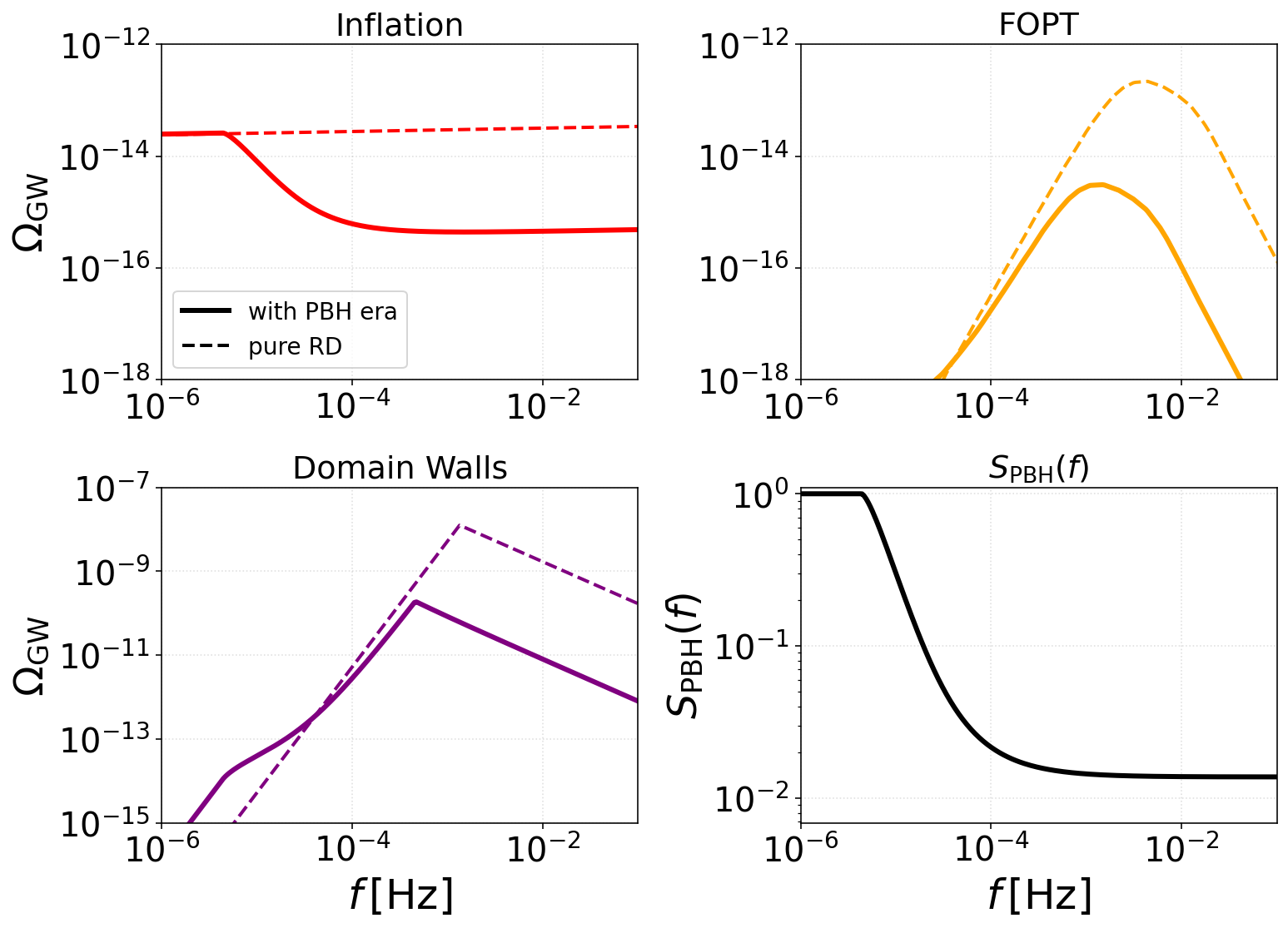} 
\caption{\it Benchmark GWBs with (solid lines) and without (dashed lines) an early PBH-dominated era. Results are shown for inflationary source (red), first-order phase transitions source (orange), and domain wall source (purple). In each case clear features and the observable frequencies can be seen. The fourth panel shows the function S defined in Eq~\ref{eq:S}. PBH parameters are $\beta_i=5\times 10^{-10}$ and $M_{PBH}=2\times10^{5} \rm g$ Further discussion is given in the text.}
    \label{fig:bench1}
\end{figure}

The quantity of interest is the gravitational wave background spectrum, defined as the fractional energy density in GWs per logarithmic frequency interval,
\begin{equation}
    \Omega_{\rm GW}(a,f)
    \equiv
    \frac{1}{\rho_{\rm tot}(a)}
    \frac{d\rho_{\rm GW}(a)}{d\ln f}.
\end{equation}
For the moment we shall assume that there is a gravitational wave background that is no longer being sourced and is freely propagating so that the only change is the cosmological evolution due to the change of the equation of state of the universe.\\
Each observed frequency corresponds to a particular GW mode. This mode does not immediately behave as a radiation component after production. While its physical wavelength is larger than the Hubble radius (i.e. the mode is outside the horizon), the tensor perturbation is effectively frozen and does not propagate as a wave. As the universe expands, the Hubble radius grows relative to the wavelength, and the mode eventually enters the horizon when its physical scale becomes comparable to the Hubble scale. From this point onward the mode oscillates and propagates freely, redshifting like radiation. Therefore, the evolution of the GWB begins only after horizon entry, and $\Omega_{GW}$ evolves only once the corresponding mode is inside the horizon defined by,
\begin{equation}
    k=a_{ent}H(a_{ent}),\quad k=2\pi f_0a_0
\end{equation}
where $f_0$ is the frequency as measured today and $a_0$ is the scale factor measured today which we calculate from our Boltzmann code assuming that between the end of the Boltzmann evolution and today there are no non-negligible entropy dumps,
\begin{equation}
    a_0=a_f\frac{T_f}{T_0}\left(\frac{g_{*s}(T_f)}{g_{*s}(T_0)}\right)^{\frac{1}{3}},
    \label{eq:Tf}
\end{equation}
where $a_f$ is the final scale factor of the evolution of Boltzmann equations, $T_f$ and $T_0$ are the final temperature of Boltzmann's and the temperature today respectively, $g_*$ is the degrees of freedom for the corresponding times.\\
Once a mode has entered the horizon, it propagates freely and its subsequent evolution is determined by the background expansion history. For a freely propagating GWB, the modified spectrum can be written as \cite{Boyle_2008,Ghoshal:2026ros,Spalding:2026pmp}
\begin{equation}
    \Omega_{\rm GW}(a_f,f)
    =
    \Omega_{\rm GW}^{\rm RD}(f')\,
    C(w_{\rm ent})\,
    \exp\left[
    \int_{\ln a_{\rm ent}(f)}^{\ln a_f}
    \left(3w(a)-1\right)\,d\ln a
    \right],
    \qquad
    f'=f\,\Delta^{1/3},
    \label{eq:MDeffect}
\end{equation}
where $a_{\rm ent}(f)$ is the scale factor at which the mode of observed frequency $f$ re-enters the horizon, defined by $k=a_{\rm ent}H(a_{\rm ent})$, and $w_{\rm ent}=w(a_{\rm ent})$ is the equation-of-state parameter at horizon entry. The factor $C(w_{\rm ent})$ accounts for the evolution of the tensor mode through horizon crossing and is given by \cite{Boyle_2008}
\begin{equation}
    C(w[k=aH])
    =
    \frac{\Gamma^2\!\left(\alpha+\tfrac{1}{2}\right)}{\pi}
    \left(\frac{2}{\alpha}\right)^{2\alpha},
    \qquad
    \alpha=\frac{2}{1+3w}.
    \label{eq:C(w)}
\end{equation}
This expression assumes that $w$ varies slowly during horizon crossing. It gives $C=1$ for RD and $C=9/16$ for MD, corresponding to an $\mathcal{O}(1)$ correction to the amplitude.\\
The exponential factor in Eq.~\eqref{eq:MDeffect} captures the subsequent redshifting of the GW energy density relative to the total background. During RD, $w=1/3$ and no additional suppression is generated, whereas during an eMD phase, $w<1/3$, so modes already inside the horizon are suppressed.\\
The frequency rescaling $f'=f\Delta^{1/3}$ arises from the entropy injection associated with PBH evaporation. The resulting increase in the radiation temperature modifies the relation between scale factor and temperature relative to the adiabatic RD case, shifting the correspondence between the source frequency and the frequency observed today.\\ 
A useful quantity for identifying the frequencies at which deviations from RD occur is defined as \cite{Spalding:2026pmp}
\begin{equation}
    S(f)=C(w_{\mathrm{ent}})\exp\left[\int_{\ln a_{\mathrm{ent}}(f)}^{\ln a_f}(3w(a)-1)\, d\ln a \right]\ .\label{eq:S}
\end{equation}
This expression shows that modes entering the horizon after the period of eMD experience no suppression, since \(S=1\) in this regime. In contrast, modes that enter during or before the eMD era are suppressed, with modes spending a longer time in the early matter phase undergoing greater dilution.\\
To demonstrate the broad applicability of this framework, we consider several benchmark gravitational wave backgrounds generated during RD, followed by a period of eMD driven by PBH domination for inflationary \cite{Caprini_2016,Caprini_2018,Caldwell:2018giq,Boyle_2008,Boyle:2005se,Barman:2023ktz}, domain wall \cite{Vilenkin:1981zs,Vachaspati:1984gt,Blanco_Pillado_2017,Hiramatsu_2010,Barman:2022yos,Barman:2023fad}, first-order phase transition \cite{Caprini_2016,Hindmarsh:2015qta,Crawford:2024nun,Barman:2026kab} GWBs in Fig \ref{fig:bench1}. 
\\
In each scenario, a distinct signature emerges through deviations from pure RD. Although the detailed form of the modified spectrum depends on the underlying source, the effect of the PBH-dominated era is universal: modes that enter the horizon after the early matter-dominated phase are unaffected, whereas modes that enter before or during this epoch experience a suppression relative to the RD prediction. For an underlying spectrum scaling locally as $\Omega_{\rm GW}\propto f^n$, the matter-dominated evolution introduces an additional $f^{-2}$ dependence for modes entering during this period, giving rise to a characteristic $f^{n-2}$ segment. The subsequent entropy injection from PBH evaporation additionally shifts the spectrum in frequency. \\
The imprint can be characterised by two frequencies associated with the beginning and end of the eMD era. We denote by $f_1$ the characteristic frequency associated with the end of PBH domination and the return to RD, while $f_2$ corresponds to its onset. Modes with $f\lesssim f_1$ enter the horizon after the PBH-dominated era and therefore retain the RD evolution. Modes with $f_1\lesssim f\lesssim f_2$ enter during the eMD phase and acquire the characteristic change in spectral slope, while modes with $f\gtrsim f_2$ entered before PBH domination and experience the full suppression accumulated throughout the MD epoch. These two frequencies therefore encode the duration and location of the PBH-dominated era in the GW spectrum and form the basis of the reconstruction developed in the following subsection.
\subsection{Mapping Observables to PBH Parameters}
\begin{figure}[h!]
\centering
\includegraphics[width=0.7\linewidth]{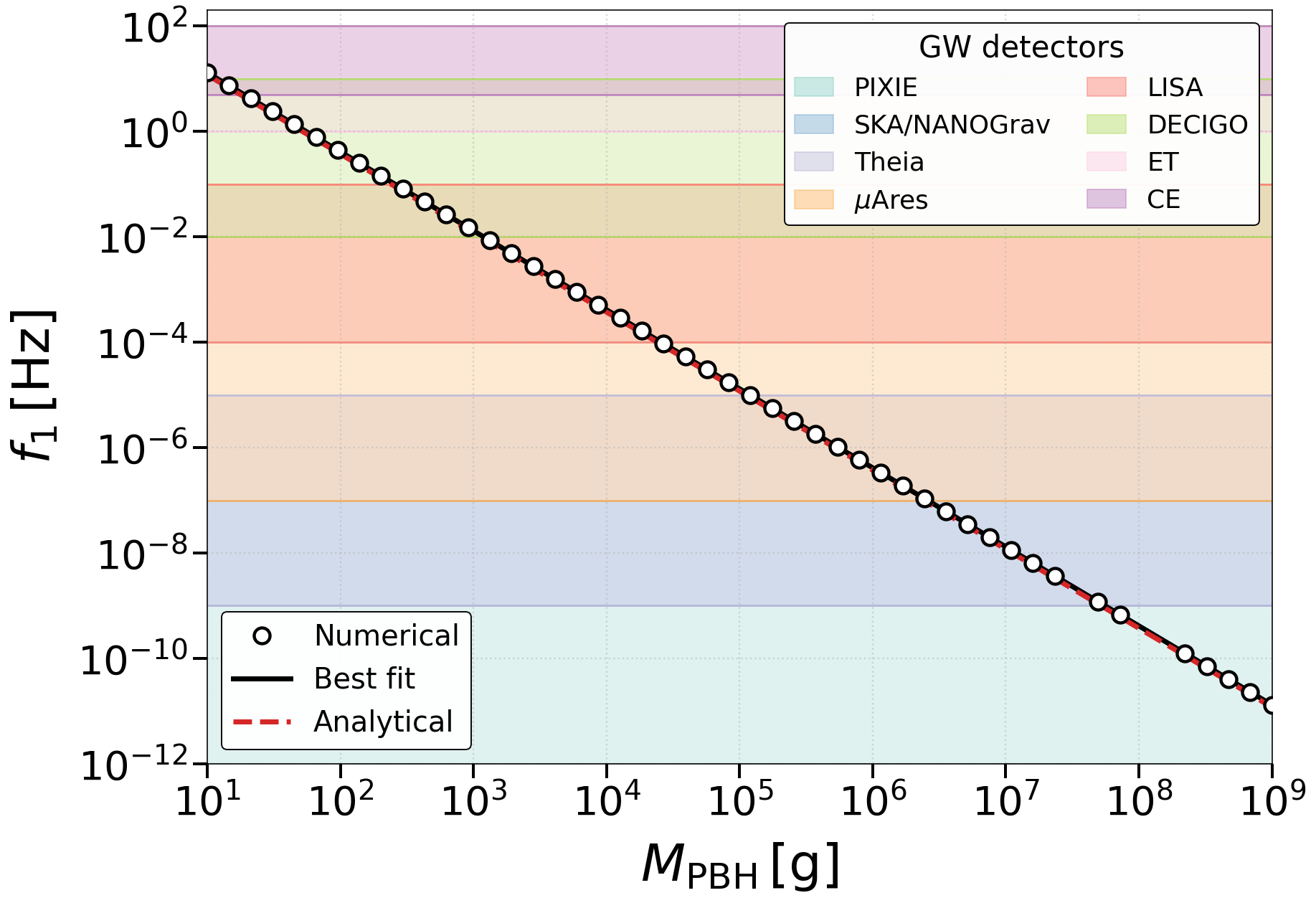} 
\caption{\it Characteristic frequency $f_1$ associated with the end of the PBH-dominated era as a function of the PBH mass. The numerical results are shown by the white circles, together with the numerical best fit (solid black) and the analytical approximation (red dashed). The shaded regions indicate the characteristic frequency ranges of current and proposed GW experiments. Further discussion is given in the text.}
    \label{fig:f1scan}
\end{figure}

\begin{figure}[h!]
\centering
\includegraphics[width=1\linewidth]{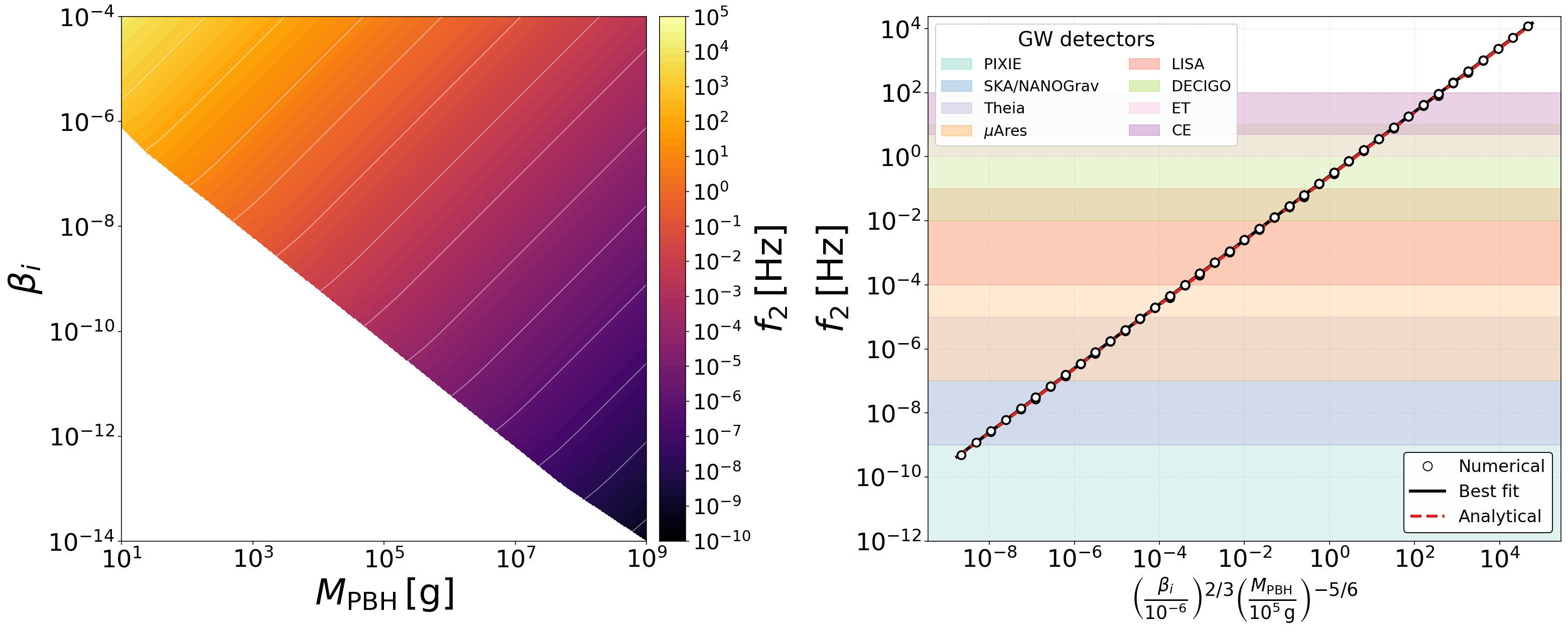} 
\caption{\it Characteristic frequency $f_2$, associated with the onset of the PBH-dominated era. \textbf{Left panel:} shows the numerical values of $f_2$ across the $(M_{\rm PBH},\beta_i)$ parameter space, restricted to the region in which PBHs dominate before evaporating. \textbf{Right  panel:} shows the same results as a function of the combination $\beta_i^{2/3}M_{\rm PBH}^{-5/6}$, demonstrating the collapse onto a single scaling relation. Numerical results are shown by the white circles, together with the numerical best fit (solid black) and analytical approximation (red dashed). The shaded regions indicate the characteristic frequency ranges of current and proposed GW experiments. Further discussion is given in the text.}
    \label{fig:f2scan}
\end{figure}

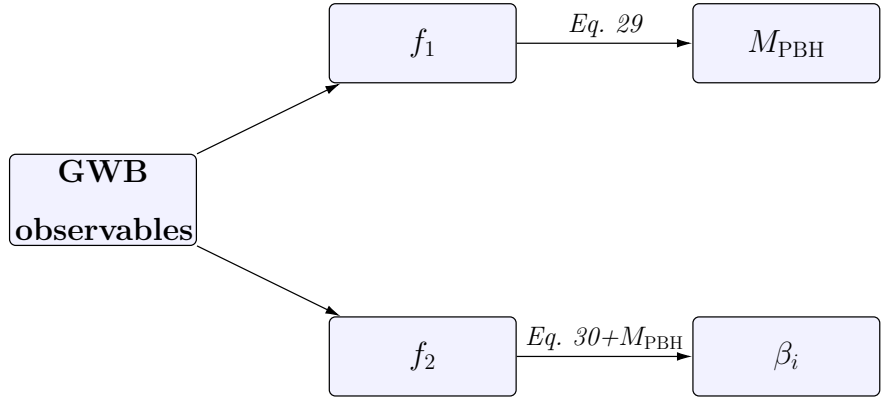
\begin{figure}[h!]
    \centering
    \resizebox{0.7\linewidth}{!}{%
    \begin{tikzpicture}[
        node distance=3.0cm and 2.6cm,
        box/.style={
            rectangle, rounded corners, draw=black, fill=blue!5, thick,
            text centered, text width=4.0cm, minimum height=1.8cm,
            font=\bfseries\LARGE    
        },
        arrow/.style={thick, -{Latex[length=3.5mm,width=2mm]}},
        eqnlabel/.style={midway, above, sloped, font=\Large\itshape},
        eqnlabelV/.style={pos=0.5, left, font=\Large\itshape}
    ]

    \node[box] (gwb) {GWB\\ observables};

    \node[box, above right=1.6cm and 3.cm of gwb] (fbrk) {$f_1$};
    \node[box, below right=1.6cm and 3.cm of gwb] (fdom) {$f_2$};

    \node[box, right=4cm of fdom] (tdom) {$\beta_i$};
    \node[box, right=4cm of fbrk] (tend) {$M_{\rm PBH}$};

    \draw[arrow] (gwb) -- (fbrk);
    \draw[arrow] (gwb) -- (fdom);

    \draw[arrow] (fdom) -- node[eqnlabel] {Eq.~\ref{eq:f2}+$M_{\rm PBH}$} (tdom);
    \draw[arrow] (fbrk) -- node[eqnlabel] {Eq.~\ref{eq:f1}} (tend);

    \end{tikzpicture}%
    }

    \caption{
        \it Schematic illustrating the connection between a GWB (GWB) and PBH parameters. A stochastic GWB can contain two characteristic frequencies: $f_2$, marking the onset of an eMD era, and $f_1$, corresponding to the transition back to RD. From $f_1$, the PBH mass can be inferred via Eq.~\ref{eq:f1}. Meanwhile, $f_2$ determines the combination of the PBH mass and initial abundance through Eq.~\ref{eq:f2}. Combining this with the value of $M_{\rm PBH}$ extracted from $f_1$ allows the initial abundance to be determined directly, thereby establishing a direct correspondence between GW observables and the PBH parameters.
  }
    \label{fig:GWB_inference_chain}
\end{figure}

We define the first breaking frequency, $f_1$ as the point where S drops below $0.95$ as a way of measuring the end of eMD \footnote{The precise threshold used to define $f_1$ is somewhat arbitrary. We adopt a $5\%$ deviation, $S(f_1)=0.95$, as a practical criterion that is large enough to identify the onset of a meaningful departure from RD evolution while remaining close to the end of the eMD era.}. We scan over PBH parameters and for each evolution extract this value, map to a modern day frequency and record the value of $f_1$. The scan for $f_1$ yields the simple best-fit relation
\begin{equation}
    f_1 =  390\left(\frac{1\,\mathrm{g}}{M_{\rm PBH}}\right)^{3/2}
    \text{Hz},
    \label{eq:f1}
\end{equation}
There is no dependence on the initial PBH abundance, since this characteristic frequency corresponds to the end of the eMD era, which is set by the PBH evaporation rate and therefore depends only on the PBH mass. In Sec.~\ref{appendix:f1}, we derive the corresponding analytic approximation and recover the same mass scaling, with a slightly different proportionality constant. The numerical scan, together with the frequency ranges probed by current and upcoming GW experiments, is shown in Fig.~\ref{fig:f1scan}.
\\
\noindent The full PBH mass range capable of producing an eMD era before BBN, $10\,{\rm g}\lesssim M_{\rm PBH}\lesssim10^9\,{\rm g}$, maps onto frequencies covered by current and upcoming GW experiments. We stress, however, that lying within the nominal frequency band of a given experiment does not by itself guarantee detectability: the amplitude of the underlying GWB must also exceed the corresponding experimental sensitivity.\\
At the lowest frequencies, the nanohertz signal reported by NANOGrav is particularly interesting. If this signal is confirmed to be primordial in origin and contains the characteristic imprint of an eMD era, its frequency range corresponds in our mapping to dominating PBHs with masses of approximately $4$--$90\,{\rm Mg}$. Thus, existing PTA observations may already be probing part of the PBH parameter space considered here. While such an imprint would not uniquely identify a PBH-dominated phase, it would provide a valuable constraint on the class of scenarios capable of producing an eMD era.
It has been shown that the characteristic frequency $f_1$ remains well defined even for GWBs that continue to be sourced during the eMD era \cite{Ghoshal:2026ros,Spalding:2026pmp}. Consequently, the absence of the corresponding spectral feature in an observed GWB can already be used to exclude regions of PBH parameter space for which an early PBH-dominated phase would otherwise predict such a feature.\\
For a freely propagating GWB, additional information can be extracted from a second characteristic frequency, $f_2$, associated with the onset of the eMD era. We define $f_2$ as the frequency at which the suppression function satisfies $S(f_2)=1.05\,S_{\min}$ and perform an analogous scan for the second characteristic frequency \footnote{As for the definition of $f_1$, this threshold is somewhat arbitrary. We choose a $5\%$ offset from the asymptotic values so that $f_1$ and $f_2$ are defined consistently.}. Importantly, changing the threshold changes the numerical normalisation rather than the underlying scaling with $M_{\rm PBH}$ and $\beta_i$. If the unmodified source spectrum can be reconstructed independently, then $S(f)$ can be inferred observationally and the same operational definition can be applied. The results of the numerical scan are shown in Fig.~\ref{fig:f2scan}.
The numerical scan over the allowed $(M_{\rm PBH},\beta_i)$ parameter space shows that all extracted values of $f_2$ collapse onto a single power law. The fitted exponent is essentially unity, confirming the scaling $f_2\propto\beta_i^{2/3}M_{\rm PBH}^{-5/6}$:

\begin{equation}
f_2= 0.250{\rm Hz}
\left(\frac{\beta_i}{10^{-6}}\right)^{2/3}
\left(\frac{M_{\rm PBH}}{10^5\rm g}\right)^{-5/6}\ .
\label{eq:f2}
\end{equation}
The dependence of $f_2$ has a simple physical interpretation. Increasing the initial PBH abundance causes the PBHs to overtake the radiation density earlier, shifting the onset of the eMD era to earlier times and therefore to larger characteristic frequencies. Similarly, for fixed $\beta_i$, decreasing the PBH mass shifts the onset of PBH domination to earlier times, again increasing $f_2$. Conversely, larger PBH masses and smaller initial abundances delay the onset of PBH domination and move the corresponding feature to lower frequencies.\\
As for $f_1$, we derive an analytic approximation for $f_2$ in Sec.~\ref{appendix:f2}. The numerical result reproduces the same scaling, $f_2\propto\beta_i^{2/3}M_{\rm PBH}^{-5/6}$, providing a direct validation of the analytic dependence.\\
We emphasise, however, that unlike $f_1$, the relation for $f_2$ can only be used when the GWB is freely propagating during the eMD phase, since continued sourcing can obscure or modify the feature associated with the onset of PBH domination. As with $f_1$, the corresponding part of the GWB must also lie above the experimental sensitivity in order for the feature to be observable.\\
Together with Eq.~\ref{eq:f1}, Eq.~\ref{eq:f2} constitutes one of the key results of this work. While $f_1$ determines the PBH mass through the end of the PBH-dominated era, $f_2$ probes the combination of mass and initial abundance that determines its onset. A measurement of both characteristic frequencies therefore provides a direct mapping from the GW spectrum to the underlying PBH mass and initial abundance.\\
These relations provide a one-to-one mapping between the observable frequencies of the GWB and the decay rate and product of the mass and initial abundance. We show this as a schematic in Fig \ref{fig:GWB_inference_chain}.

\subsection{Induced GWs from PBH dominance}

Another important feature of a PBH-dominated era is the generation of an additional stochastic GWB. The discrete nature of the PBH population gives rise to Poisson fluctuations in the PBH density \cite{Papanikolaou:2020qtd,Domenech:2023jve,Domenech:2024wao,Domenech:2020ssp,Papanikolaou:2022chm,Papanikolaou:2024kjb,He:2024luf,Gouttenoire:2026pem,Gouttenoire:2026pmp}, while the primordial adiabatic scalar perturbations are also enhanced by the transition from PBH domination back to RD. These scalar perturbations source tensor modes at second order \cite{Domenech:2021ztg,Baumann:2007,Ananda:2006,Assadullahi:2009,del-Corral:2025fzz}, producing two distinct contributions,
\begin{equation}
    \Omega_{\rm GW}^{\rm PBH}(f)
    =
    \Omega_{\rm GW}^{\rm ad}(f)
    +
    \Omega_{\rm GW}^{\rm iso}(f),
\end{equation}
where the lower-frequency peak is sourced by the primordial adiabatic perturbations and the higher-frequency peak by the PBH Poisson/isocurvature fluctuations. The resulting doubly peaked spectrum and analytic approximations are derived in Ref.~\cite{Bhaumik:2022pil}.

For non-spinning PBHs, the dominant branches approaching the two resonances may be written schematically as
\begin{equation}
    \Omega_{\rm GW}^{\rm ad}(f)
    \simeq
    \Omega_{\rm ad}^{\rm pk}
    \left(\frac{f}{f_m}\right)^7,
    \qquad
    \Omega_{\rm GW}^{\rm iso}(f)
    \simeq
    2.25\,\Omega_{\rm iso}^{\rm pk}
    \left(\frac{f}{f_{\rm UV}}\right)^{11/3},
\end{equation}
with
\begin{align}
    f_m &\simeq
    2.36\times10^{6}\,{\rm Hz}\,
    \beta_i^{2/3}
    \left(\frac{M_{\rm PBH}}{\rm g}\right)^{-5/6},
    \\
    f_{\rm UV} &\simeq
    9.7\times10^{2}\,{\rm Hz}
    \left(\frac{M_{\rm PBH}}{10^{4}\,{\rm g}}\right)^{-5/6},
    \\
    \Omega_{\rm ad}^{\rm pk}
    &\simeq
    6.9\times10^{-6}\,
    \beta_i^{14/3}
    \left(\frac{M_{\rm PBH}}{\rm g}\right)^{14/3},
    \\
    \Omega_{\rm iso}^{\rm pk}
    &\simeq
    1.9\times10^{6}\,
    \beta_i^{16/3}
    \left(\frac{M_{\rm PBH}}{\rm g}\right)^{34/9}.
\end{align}
The full analytic result contains additional $f^3$ and $f$ branches below the first resonance and the resonant caps of both peaks; these full piecewise expressions from Ref.~\cite{Bhaumik:2022pil} are used below. The peak amplitude may also be related directly to the PBH scales, as discussed in Ref.~\cite{del-Corral:2025fca}.

\begin{figure}[H]
    \centering
    \includegraphics[width=\linewidth]{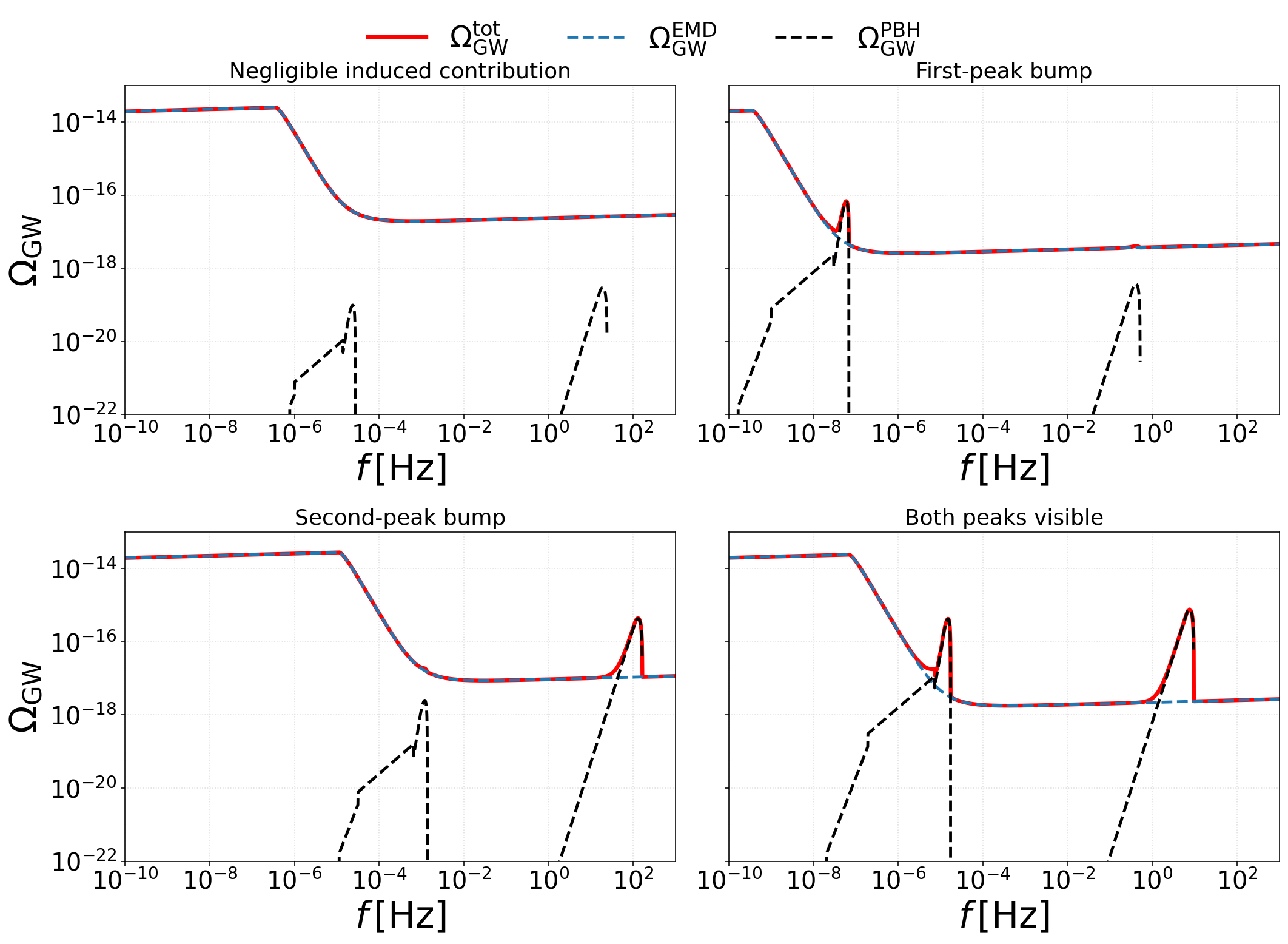}
    \caption{\it Total gravitational wave background (red), the EMD modified inflationary background (blue dashed), and PBH-induced background (black dashed). From upper left to lower right the parameters are, $(\beta_i,M_{\rm PBH})=(10^{-9},10^{6}\,{\rm g})$, $(3.2\times10^{-11},10^{8}\,{\rm g})$, $(1.5\times10^{-8},10^{5}\,{\rm g})$, and $(10^{-9},3\times10^{6}\,{\rm g})$.}
    \label{fig:PBH_GWB}
\end{figure}

This induced contribution is distinct from the propagation-induced modification of a pre-existing GWB considered in Secs.~IV\,A and IV\,B. The observable spectrum is therefore
\begin{equation}
    \Omega_{\rm GW}^{\rm tot}(f)
    =
    \Omega_{\rm GW}^{\rm EMD}(f)
    +
    \Omega_{\rm GW}^{\rm PBH}(f).
    \label{eq:Omega_GW_total}
\end{equation}
Depending on the amplitude of the pre-existing GWB relative to the two PBH-induced peaks, the induced contribution can either remain negligible, furnish additional information and an independent consistency check of the PBH-dominated scenario, or become large enough to blur the spectral feature at $f_2$ used in the reconstruction above.\\
To illustrate when the induced component affects the reconstruction, we consider four representative inflationary benchmarks in Fig.~\ref{fig:PBH_GWB}: a case in which the induced GWB is negligible, cases in which either the adiabatic or Poisson peak becomes visible, and a case in which both peaks modify the total spectrum.

Figure~\ref{fig:PBH_GWB} shows that an observable PBH-induced contribution does not necessarily erase the propagation-induced eMD signature. When the induced signal is small, the total spectrum is essentially identical to $\Omega_{\rm GW}^{\rm EMD}$. As its amplitude increases, either induced peak can appear as an additional local feature while the broad eMD suppression remains identifiable. There is also a regime in which both peaks are visible without destroying the underlying propagation signature. Only once the lower-frequency induced peak becomes sufficiently large to obscure the feature associated with $f_2$ would the simple reconstruction of Sec.~IV\,B need to be replaced by a joint fit to the complete spectrum. We have the generic hierarchy
\begin{equation}
    f_1 < f_m< f_2< f_{\rm UV}.
\end{equation}
The induced high-frequency peak can therefore provide an additional observable while leaving the two propagation-induced frequencies used in the reconstruction accessible.

This leads to a further consistency test of the scenario. The PBH cosmology is specified by only two parameters, $(M_{\rm PBH},\beta_i)$, while the complete GW spectrum may potentially contain four characteristic frequency scales,
\begin{equation}
    \left\{f_1,\;f_2,\; f_m,\; f_{\rm UV}\right\}.
\end{equation}
As shown above, $f_1$ determines $M_{\rm PBH}$, while $f_2$ subsequently determines $\beta_i$. Then the analysis of \cite{Bhaumik:2022pil} predicts the location of the PBH-induced peaks. A simultaneous observation of three of even all four frequencies would therefore over-constrain the two-parameter PBH scenario and provide a non-trivial consistency test of an early PBH-dominated epoch.

\section{Discussion \& Conclusion}\label{sec:conclusion}
In this work, we have investigated the possibility of probing an eMD era sourced by primordial black holes through its imprint on a stochastic GWB. PBHs satisfying Eqn.~\ref{eq:PBH_dom_condition} that evaporate before BBN temporarily dominate the energy density of the Universe, modifying the expansion history over scales that are otherwise extremely difficult to access observationally. Since GWs respond directly to the background equation of state after horizon entry, such a PBH-dominated phase leaves a characteristic suppression in the GW spectrum.

The resulting modification of a GWB is characterised by two frequencies. The first, $f_1$, probes the end of PBH domination and the subsequent return to RD. Since this scale is controlled primarily by the PBH lifetime, it depends only on the PBH mass. As shown in Eq.~\ref{eq:f1}, the numerical result follows the analytic scaling $f_1\propto M_{\rm PBH}^{-3/2}$ to good accuracy. Consequently, a measurement of $f_1$ provides a direct determination of $M_{\rm PBH}$. This relation remains valid even while the GWB is still being sourced, provided that the underlying unmodified spectrum can be independently reconstructed. For GWBs such as inflationary backgrounds \cite{Caprini_2016,Caprini_2018,Caldwell:2018giq,Boyle_2008,Boyle:2005se,Barman:2023ktz}, where the primordial spectral shape is well understood, this can be done straightforwardly. For other sources, such as domain walls \cite{Vilenkin:1981zs,Vachaspati:1984gt,Blanco_Pillado_2017,Hiramatsu_2010,Barman:2022yos,Barman:2023fad} or first-order phase transitions \cite{Caprini_2016,Hindmarsh:2015qta,Crawford:2024nun,Barman:2026kab}, intrinsic features in the spectrum may partially or completely obscure the imprint of the eMD era.

For a freely propagating GWB, additional information can be extracted from a second characteristic frequency, $f_2$, which marks the onset of the suppression and therefore depends on both the PBH mass and its initial abundance. The numerical scan over the PBH-dominated region confirms the analytic scaling $f_2\propto\beta_i^{2/3}M_{\rm PBH}^{-5/6}$, with the resulting numerical relation given in Eq.~\ref{eq:f2}. Once $M_{\rm PBH}$ has been determined independently from $f_1$, the measurement of $f_2$ therefore directly determines $\beta_i$. Together, Eqs.~\ref{eq:f1} and \ref{eq:f2} establish a one-to-one mapping between the two characteristic frequencies imprinted on the GW spectrum and the two underlying PBH parameters $(M_{\rm PBH},\beta_i)$ as shown in Fig \ref{fig:GWB_inference_chain}.

An important feature of this approach is that the suppression is not tied to a particular mechanism for generating the primordial GWB. Rather, it arises from the subsequent propagation of gravitational waves through the modified expansion history. The same characteristic structure can therefore be imprinted on backgrounds produced by inflation, first-order phase transitions, annihilating domain walls, or other sufficiently early and transient sources. This source independence significantly broadens the range of possible observational applications. If the GWB is still being sourced then only the relation for $f_1$ is valid and the mass can be determined while $f_2$ is then source dependent.

The wide frequency coverage of present and proposed GW experiments is particularly attractive in this context. The characteristic frequencies associated with PBHs over the mass interval
$10\,{\rm g}\lesssim M_{\rm PBH}\lesssim10^9\,{\rm g}$ extend across many orders of magnitude in frequency, overlapping the ranges targeted by pulsar timing arrays, astrometric probes, space-based interferometers and future ground-based detectors. We emphasise, however, that a characteristic frequency lying within an experimental frequency range does not by itself guarantee observability; the underlying GWB must also be sufficiently large for the spectral feature to lie above the relevant detector sensitivity. We found that the nanohertz signal reported by NANOGrav, if primordial in origin, would probe PBH masses in the range $4$--$90\,{\rm Mg}$.

A complementary signature arises from GWs induced at second order by scalar
perturbations during the PBH-dominated era. As discussed in Sec.~IV\,C, the
resulting spectrum can contain two additional peaks: a lower-frequency
adiabatic peak around $f_m$ and a higher-frequency Poisson/isocurvature peak
around $f_{\rm UV}$. Depending on $(M_{\rm PBH},\beta_i)$, either, both, or
neither of these peaks may be visible in the total GWB without necessarily
obscuring the propagation-induced features at $f_1$ and $f_2$. The complete
spectrum may therefore contain up to four characteristic frequency scales,
$f_1$, $f_2$, $f_m$, and $f_{\rm UV}$. Since the underlying PBH cosmology is
specified by only two parameters, the observation of additional induced
features would overconstrain the scenario and provide a non-trivial
consistency test of an early PBH-dominated epoch. In particular, once
$M_{\rm PBH}$ and $\beta_i$ are reconstructed from $f_1$ and $f_2$, the
locations and amplitudes of the induced peaks are no longer independent.
For sufficiently large induced amplitudes, however, the lower-frequency
peak can distort the feature associated with $f_2$, in which case the
propagation-induced and PBH-induced components must be analysed jointly.

The main result of this work is therefore a direct correspondence between otherwise inaccessible PBH parameters and observable spectral features in gravitational waves. PBHs that disappear long before BBN can leave a surviving record through the non-standard expansion history they generate. If a primordial stochastic GWB is observed over a sufficiently broad frequency range, the two characteristic frequencies $f_1$ and $f_2$ could therefore provide a direct measurement of both the PBH mass and its initial abundance, opening a GW window onto an early PBH-dominated Universe.

\section*{Acknowledgements}
The authors thank Lucien Heurtier for insightful discussions on PBH domination and Anish Ghoshal for valuable comments on the first version of the manuscript and discussions on the PBH-induced gravitational-wave background. AS acknowledges the STFC Consolidated Grant ST/X000583/1 and thanks the University of Southampton School of Physics and Astronomy for the support of a Mayflower PhD scholarship. D.dC acknowledges the support of the grant No. 2021/42/E/ST9/00260 from the National
Science Centre, Poland.
\appendix
\section{Analytic Approximations of Key quantities}
\label{appendix}
In this appendix, we derive analytic approximations for the key quantities used throughout this work: the condition for PBH domination, the associated entropy injection, and the two characteristic GW frequencies, $f_1$ and $f_2$. These results provide simple physical interpretations of the scalings observed in the full numerical analysis.

\subsection{Condition for PBH domination}
\label{appendix:PBHdom}

The boundary for PBH domination can be understood analytically by neglecting the energy injected into radiation during the early stages of evaporation. Prior to PBH domination, the Universe therefore remains approximately RD, such that
\begin{equation}
    a(t) \simeq \left(1+2H_i t\right)^{1/2}.
\end{equation}
The ratio of the PBH and radiation energy densities then evolves as
\begin{equation}
    \frac{\rho_{\rm PBH}}{\rho_R}
    \simeq
    \beta_i
    \left(1+2Ay\right)^{1/2}
    \left(1-y\right)^{1/3},
    \qquad
    y\equiv\frac{t}{\tau_{\rm PBH}},
\end{equation}
where
\begin{equation}
    A\equiv H_i\tau_{\rm PBH}
    =
    \frac{5120\pi\gamma}{g_H}
    \left(\frac{M_{\rm PBH}}{M_{\rm pl}}\right)^2,
\end{equation}
and $M_{\rm pl}=G^{-1/2}$ denotes the unreduced Planck mass. The maximum of this ratio occurs at
\begin{equation}
    y_{\rm max}=\frac{3A-1}{5A},
\end{equation}
and requiring $\max(\rho_{\rm PBH}/\rho_R)=1$ gives the critical initial abundance
\begin{equation}
    \beta_{\rm crit}^{\rm ana}
    =
    \sqrt{\frac{5}{3}}\,
    \frac{(5A)^{1/3}}{(1+2A)^{5/6}}.
\end{equation}
For the parameter space considered in this work, $A\gg1$, and the critical abundance simplifies to
\begin{equation}
    \beta_{\rm crit}^{\rm ana}
    \simeq
    1.24\left(H_i\tau_{\rm PBH}\right)^{-1/2}
    \simeq
    4.91\times10^{-6}
    \left(\frac{M_{\rm PBH}}{\rm g}\right)^{-1},
\end{equation}
where we have used $g_H=106.75$ and $\gamma=0.2$. This reproduces the scaling $\beta_{\rm crit}\propto M_{\rm PBH}^{-1}$ found in the full numerical evolution, with the small difference in normalisation arising from the approximations made above.

\subsection{Entropy Injection Factor}
\label{appendix:entropy}

An analytic estimate of the entropy injection can be obtained in the sudden-evaporation approximation. Prior to significant PBH evaporation, the PBH and radiation energy densities scale as $\rho_{\rm PBH}\propto a^{-3}$ and $\rho_R\propto a^{-4}$, respectively. Their ratio therefore evolves as
\begin{equation}
    \frac{\rho_{\rm PBH}}{\rho_R}
    \simeq
    \beta_i\frac{a}{a_i},
\end{equation}
where we have assumed $\beta_i\ll1$. PBH domination begins when $\rho_{\rm PBH}=\rho_R$, giving
\begin{equation}
    \frac{a_{\rm dom}}{a_i}
    \simeq
    \beta_i^{-1}.
\end{equation}
At this time, $\rho_{\rm tot}=2\rho_R$. Since the radiation energy
density evolves as $\rho_R\propto a^{-4}$ prior to domination, the
Friedmann equation gives
\begin{equation}
    H_{\rm dom}^2
    \simeq
    2H_i^2
    \left(\frac{a_i}{a_{\rm dom}}\right)^4,
\end{equation}
and hence
\begin{equation}
    H_{\rm dom}
    \simeq
    \sqrt{2}\,H_i\beta_i^2.
\end{equation}
During PBH domination, the expansion is approximately MD,
so that
\begin{equation}
    H(a)
    \simeq
    H_{\rm dom}
    \left(\frac{a_{\rm dom}}{a}\right)^{3/2}.
\end{equation}
Integrating the Friedmann equation from the onset of PBH domination gives
\begin{equation}
    \frac{a(t)}{a_{\rm dom}}
    =
    \left[
        1+\frac{3}{2}H_{\rm dom}(t-t_{\rm dom})
    \right]^{2/3}.
\end{equation}
For a sufficiently long PBH-dominated era,
$t_{\rm ev}\simeq\tau_{\rm PBH}\gg t_{\rm dom}$ and
$H_{\rm dom}\tau_{\rm PBH}\gg1$, such that
\begin{equation}
    \frac{a_{\rm ev}}{a_{\rm dom}}
    \simeq
    \left(
        \frac{3}{2}H_{\rm dom}\tau_{\rm PBH}
    \right)^{2/3}.
\end{equation}
where $\tau_{\rm PBH}$ denotes the PBH lifetime. The ratio of the PBH and radiation energy densities immediately before evaporation is consequently
\begin{equation}
    \left.
    \frac{\rho_{\rm PBH}}{\rho_R}
    \right|_{\rm ev}
    \simeq
    \left(
        \frac{3\sqrt{2}}{2}
        H_i\tau_{\rm PBH}\beta_i^2
    \right)^{2/3}.
\end{equation}
For a sufficiently long PBH-dominated era, the pre-existing radiation is negligible compared with the energy density stored in the PBHs. In the sudden-evaporation approximation, this energy is converted instantaneously into radiation, such that the entropy injection factor becomes
\begin{equation}
    \Delta
    \simeq
    \left(
        \left.
        \frac{\rho_{\rm PBH}}{\rho_R}
        \right|_{\rm ev}
    \right)^{3/4}
    =
    \beta_i
    \left(
        \frac{3\sqrt{2}}{2}
        H_i\tau_{\rm PBH}
    \right)^{1/2}.
    \label{eq:Delta_analytic}
\end{equation}
Using the unreduced Planck mass $M_{\rm pl}=G^{-1/2}$, the initial Hubble rate and PBH lifetime are
\begin{equation}
    H_i
    =
    \frac{\gamma M_{\rm pl}^2}{2M_{\rm PBH}},
    \qquad
    \tau_{\rm PBH}
    =
    \frac{10240\pi}{g_H}
    \frac{M_{\rm PBH}^3}{M_{\rm pl}^4}.
\end{equation}
It follows that
\begin{equation}
    \Delta_{\rm ana}
    \simeq
    \beta_i
    \sqrt{
        \frac{3\sqrt{2}}{2}
        \frac{5120\pi\gamma}{g_H}
    }
    \frac{M_{\rm PBH}}{M_{\rm pl}}.
\end{equation}
For $\gamma=0.2$ and $g_H=106.75$, this gives
\begin{equation}
    \Delta_{\rm ana}
    \simeq
    3.67\times10^{5}\,
    \beta_i
    \left(
        \frac{M_{\rm PBH}}{\mathrm{g}}
    \right).
    \label{eq:Delta_analytic_numeric}
\end{equation}
Thus, within the sudden-evaporation approximation, the entropy injection depends only on the combination $\beta_i M_{\rm PBH}$, reproducing the linear scaling found in the full numerical evolution. The difference in normalisation arises from the assumption of instantaneous evaporation, since in the full evolution part of the PBH energy density is transferred to radiation at earlier times and subsequently redshifts before evaporation is complete. 
\subsection{First Characteristic Frequency $f_1$}
\label{appendix:f1}

The first characteristic frequency, $f_1$, is associated with the end of
the PBH-dominated era and the subsequent return to RD.
Analytically, we estimate this frequency by identifying it with the
comoving Hubble scale at the time of PBH evaporation. We therefore first
determine the Hubble scale and radiation temperature at evaporation, and
then redshift the corresponding mode to the present day.

We begin by recalling the PBH lifetime
\begin{equation}
    \tau_{\rm PBH}
    =
    \frac{M_{\rm PBH}^3}{3\alpha}
    =
    \frac{10240\pi}{g_H}
    \frac{M_{\rm PBH}^3}{M_{\rm pl}^4}.
    \label{eq:tau_f1_analytic}
\end{equation}
For a sufficiently long PBH-dominated era, the expansion immediately
before evaporation is approximately MD, $a\propto t^{2/3}$,
so that $ H(t)\simeq\frac{2}{3t}.$ Taking the evaporation time to be $t_{\rm ev}\simeq\tau_{\rm PBH}$ therefore
gives
\begin{equation}
    H_{\rm ev}
    \simeq
    \frac{2}{3\tau_{\rm PBH}}
    =
    \frac{g_H}{15360\pi}
    \frac{M_{\rm pl}^4}{M_{\rm PBH}^3},
    \label{eq:Hev_analytic}
\end{equation}
In the sudden-evaporation approximation, the PBH energy density is
converted into radiation at $t_{\rm ev}$. Immediately after evaporation,
the total energy density is therefore RD. Using the
Friedmann equation and equating to the formula for the radiation energy density,
\begin{equation}
    \rho_{\rm ev}
    =
    \frac{3M_{\rm pl}^2H_{\rm ev}^2}{8\pi}, \qquad \rho_R
    =
    \frac{\pi^2}{30}g_{\ast,\rm ev}T_{\rm ev}^4,
\end{equation}
gives the radiation temperature immediately after evaporation,
\begin{align}
    T_{\rm ev}
    &=
    \left(
        \frac{90}{8\pi^3g_{\ast,\rm ev}}
        M_{\rm pl}^2H_{\rm ev}^2
    \right)^{1/4}
    \nonumber\\
    &\simeq
    \left(
        \frac{5M_{\rm pl}^2}
        {\pi^3g_{\ast,\rm ev}\tau_{\rm PBH}^2}
    \right)^{1/4}.
    \label{eq:Tev_analytic}
\end{align}
We now identify the first characteristic mode with the comoving Hubble
scale at evaporation, $k_1\simeq a_{\rm ev}H_{\rm ev}$. Its physical frequency observed today is therefore
\begin{equation}
    f_1
    =
    \frac{k_1}{2\pi a_0}
    =
    \frac{H_{\rm ev}}{2\pi}
    \frac{a_{\rm ev}}{a_0}.
    \label{eq:f1_redshift}
\end{equation}
After PBH evaporation, we assume that the subsequent expansion is
adiabatic. Conservation of comoving entropy then relates the scale factor at evaporation to its present-day value,
\begin{equation}
    \frac{a_{\rm ev}}{a_0}
    =
    \frac{T_0}{T_{\rm ev}}
    \left(
        \frac{g_{\ast s,0}}
             {g_{\ast s,\rm ev}}
    \right)^{1/3}.
\end{equation}
The present-day characteristic frequency is consequently
\begin{equation}
    f_1
    =
    \frac{H_{\rm ev}}{2\pi}
    \frac{T_0}{T_{\rm ev}}
    \left(
        \frac{g_{\ast s,0}}
             {g_{\ast s,\rm ev}}
    \right)^{1/3}.
    \label{eq:f1_analytic_general}
\end{equation}

Using $H_{\rm ev}=2/(3\tau_{\rm PBH})$ and Eq.~\eqref{eq:Tev_analytic} and Eq.~\eqref{eq:tau_f1_analytic} gives
\begin{equation}
    f_1
    =
    \frac{T_0}{3\pi}
    \left(
        \frac{g_{\ast s,0}}
             {g_{\ast s,\rm ev}}
    \right)^{1/3}
    \left(
        \frac{\pi^3g_{\ast,\rm ev}}{5}
    \right)^{1/4}
    \left(
        \frac{g_H}{10240\pi}
    \right)^{1/2}
    \left(
        \frac{M_{\rm pl}}{M_{\rm PBH}}
    \right)^{3/2}.
    \label{eq:f1_analytic_explicit}
\end{equation}
Taking
$g_{\ast,\rm ev}=g_{\ast s,\rm ev}=g_H=106.75$,
$g_{\ast s,0}=3.91$, $T_0=2.7255\,{\rm K}$, and
$M_{\rm pl}=1.22\times10^{19}\,{\rm GeV}$, and converting from natural
units to Hz, gives us the analytical $f_1$.
\begin{equation}
    f_1^{\rm analytic}
    \simeq
    3.73\times10^{2}\,{\rm Hz}
    \left(
        \frac{M_{\rm PBH}}{\rm g}
    \right)^{-3/2}.
    \label{eq:f1_analytic_numeric}
\end{equation}
This reproduces the $M_{\rm PBH}^{-3/2}$ scaling obtained from the full
numerical evolution. The small difference in normalisation relative to
the numerical result arises because the analytic treatment identifies
$f_1$ directly with the horizon scale at instantaneous PBH evaporation,
whereas the full evolution accounts for continuous Hawking evaporation
and defines the transition frequency from the resulting GW
suppression function.
\subsection{Second Characteristic Frequency $f_2$}
\label{appendix:f2}

The second characteristic frequency, $f_2$, is associated with the onset
of the PBH-dominated era. In analogy with the derivation of $f_1$, we
identify this frequency with the comoving Hubble scale at PBH domination, $k_2 \simeq a_{\rm dom}H_{\rm dom}$. The corresponding frequency observed today is therefore
\begin{equation}
    f_2
    =
    \frac{k_2}{2\pi a_0}
    =
    \frac{H_{\rm dom}}{2\pi}
    \frac{a_{\rm dom}}{a_0}.
    \label{eq:f2_redshift}
\end{equation}

Unlike $f_1$, the mode associated with $f_2$ enters the horizon before
PBH evaporation. The subsequent entropy injection therefore modifies the
mapping between the scale factor at PBH domination and its present-day
value. Using the entropy injection factor $\Delta$ derived in
Sec.~\ref{appendix:entropy}, we have
\begin{equation}
    \frac{a_{\rm dom}}{a_0}
    =
    \frac{T_0}{T_{\rm dom}}
    \left(
        \frac{g_{\ast s,0}}
             {g_{\ast s,\rm dom}}
    \right)^{1/3}
    \Delta^{-1/3}.
\end{equation}
Equation~\eqref{eq:f2_redshift} consequently becomes
\begin{equation}
    f_2
    =
    \frac{H_{\rm dom}}{2\pi}
    \frac{T_0}{T_{\rm dom}}
    \left(
        \frac{g_{\ast s,0}}
             {g_{\ast s,\rm dom}}
    \right)^{1/3}
    \Delta^{-1/3}.
    \label{eq:f2_analytic_general}
\end{equation}

At the onset of PBH domination, $\rho_{\rm PBH}=\rho_R$, such that the
total energy density is $\rho_{\rm tot}=2\rho_R$. Using the Friedmann
equation with the unreduced Planck mass, $M_{\rm pl}=G^{-1/2}$,
\begin{equation}
    H_{\rm dom}^2
    =
    \frac{8\pi}{3M_{\rm pl}^2}\rho_{\rm tot}
    =
    \frac{16\pi}{3M_{\rm pl}^2}\rho_R, \qquad \rho_R
    =
    \frac{\pi^2}{30}
    g_{\ast,\rm dom}T_{\rm dom}^4,
\end{equation}
gives
\begin{equation}
    T_{\rm dom}
    =
    \left(
        \frac{45M_{\rm pl}^2H_{\rm dom}^2}
             {8\pi^3g_{\ast,\rm dom}}
    \right)^{1/4}.
\end{equation}
Substituting this into Eq.~\eqref{eq:f2_analytic_general} gives
\begin{equation}
    f_2
    =
    \frac{T_0}{2\pi}
    \left(
        \frac{8\pi^3g_{\ast,\rm dom}}{45}
    \right)^{1/4}
    \left(
        \frac{g_{\ast s,0}}
             {g_{\ast s,\rm dom}}
    \right)^{1/3}
    \left(
        \frac{H_{\rm dom}}{M_{\rm pl}}
    \right)^{1/2}
    \Delta^{-1/3}.
    \label{eq:f2_analytic}
\end{equation}

As shown in Sec.~\ref{appendix:entropy}, the Hubble rate at the onset of
PBH domination is
\begin{equation}
    H_{\rm dom}
    \simeq
    \sqrt{2}\,H_i\beta_i^2, \qquad H_i
    = \frac{\gamma M_{\rm pl}^2}{2M_{\rm PBH}}.
\end{equation}
The analytic entropy injection derived in
Sec.~\ref{appendix:entropy} scales as $\Delta_{\rm ana}\propto\beta_i M_{\rm PBH}.$
Equation~\eqref{eq:f2_analytic} therefore gives
\begin{equation}
    f_2
    \propto
    H_{\rm dom}^{1/2}\Delta^{-1/3}
    \propto
    \left(
        \beta_i^2M_{\rm PBH}^{-1}
    \right)^{1/2}
    \left(
        \beta_iM_{\rm PBH}
    \right)^{-1/3}=\beta_i^{2/3}
    M_{\rm PBH}^{-5/6},
\end{equation}
Using the numerical constants adopted in the analysis gives
\begin{equation}
    f_2^{\rm analytic}
    \simeq
    1.55\times10^{-2}\,{\rm Hz}
    \left(
        \frac{\beta_i}{10^{-6}}
    \right)^{2/3}
    \left(
        \frac{M_{\rm PBH}}{10^5\,{\rm g}}
    \right)^{-5/6}.
\end{equation}
The estimate above corresponds to a mode entering the horizon exactly at
PBH domination, $\rho_{\rm PBH}=\rho_R$. To account for the definition
used in the numerical analysis, we instead determine the mode satisfying
$S(f_2)=1.05\,S_{\min}$. Prior to significant evaporation, defining
\begin{equation}
    r \equiv \frac{\rho_{\rm PBH}}{\rho_R}
    \simeq \beta_i\frac{a}{a_i},
\end{equation}
the total equation-of-state parameter is
\begin{equation}
    w=\frac{1}{3(1+r)},
    \qquad
    3w-1=-\frac{r}{1+r}.
\end{equation}
Neglecting the small horizon-crossing correction in locating the
transition, the suppression relative to its minimum is therefore
\begin{equation}
    \frac{S(r)}{S_{\min}}
    =
    \frac{1+r}{1+\beta_i}.
\end{equation}
The $5\%$ criterion consequently gives
\begin{equation}
    r_2
    =
    1.05(1+\beta_i)-1
    \simeq 0.05,
\end{equation}
where the final expression follows for $\beta_i\ll1$. During this epoch,
\begin{equation}
    aH
    =
    a_iH_i\frac{\beta_i}{r}
    \left(\frac{1+r}{1+\beta_i}\right)^{1/2}.
\end{equation}
Comparing this with the mode entering at PBH domination, $r=1$, gives
\begin{equation}
    \frac{f_2^{5\%}}{f_2^{\rm dom}}
    =
    \frac{1}{r_2}
    \left(\frac{1+r_2}{2}\right)^{1/2}
    \simeq 14.49.
\end{equation}
Applying this correction to the estimate above yields
\begin{equation}
    f_2^{\rm analytic}
    \simeq
    2.25\times10^{-1}\,{\rm Hz}
    \left(
        \frac{\beta_i}{10^{-6}}
    \right)^{2/3}
    \left(
        \frac{M_{\rm PBH}}{10^5\,{\rm g}}
    \right)^{-5/6},
\end{equation}
which is in good agreement with the numerical result. Thus, the characteristic frequency associated with the onset of PBH
domination scales as
$f_2\propto\beta_i^{2/3}M_{\rm PBH}^{-5/6}$, in agreement with the
parameter dependence found in the full numerical evolution.
\bibliographystyle{apsrev4-1}

\bibliography{ref}
\end{document}